\documentclass{imsart}
\RequirePackage[OT1]{fontenc}
\RequirePackage{amsthm,amsmath}
\RequirePackage[numbers]{natbib}
\RequirePackage[colorlinks,citecolor=blue,urlcolor=blue]{hyperref}
\usepackage{graphicx}
\usepackage{amsmath}
\usepackage{amssymb}
\usepackage{threeparttable}

\startlocaldefs
\numberwithin{equation}{section}
\theoremstyle{plain}

\endlocaldefs

\begin{document}

\begin{frontmatter}

\title{Species Abundance Distribution and Species Accumulation Curve: A General Framework and Results}
\runtitle{SAD and SAC}

\author{\fnms{Cheuk Ting} \snm{Li}\ead[label=e1]{ctli@ie.cuhk.edu.hk}}
\address{Department of Information Engineering, The Chinese University of Hong Kong, \\ Hong Kong SAR, China\\
\printead{e1}}
\and
\author{\fnms{Kim-Hung} \snm{Li}\ead[label=e2]{khli@acrc.hk}}
\address{Asian Cities Research Centre Ltd., New Treasure Centre, San Po Kong, \\ Hong Kong SAR, China
\\
\printead{e2}}

\runauthor{C.~T. Li \& K.-H. Li}

\begin{abstract}
We build a general framework which establishes a one-to-one correspondence between species abundance distribution (SAD) and species accumulation curve (SAC).
The appearance 
rates of the species and the appearance times of individuals of each species are modeled as Poisson processes. The number of species can be finite or infinite. Hill numbers are extended to the framework. We introduce a linear derivative ratio family of models, $\mathrm{LDR}_1$, of which the ratio of the first and the second derivatives of the expected SAC is a linear function. A D1/D2 plot is proposed to detect this linear pattern in the data. By extrapolation of the curve in the D1/D2 plot, a  species richness estimator that extends Chao1 estimator is introduced. The SAD of $\mathrm{LDR}_1$ is the Engen's extended negative binomial distribution, and the SAC encompasses several popular parametric forms including the power law. Family $\mathrm{LDR}_1$ is extended in two ways: $\mathrm{LDR}_2$ which allows species with zero detection probability, and $\mathrm{RDR}_1$ where the derivative ratio is a rational function. Real data are analyzed to demonstrate the proposed methods. We also consider the scenario where we record only a few leading appearance times of each species. We show how maximum likelihood inference can be performed when only the empirical SAC is observed, and elucidate its advantages over the traditional curve-fitting method.
\end{abstract}

\begin{keyword}[class=MSC]
\kwd[Primary ]{62P10}
\kwd[; secondary ]{92D40}
\end{keyword}

\begin{keyword}
\kwd{Diagnostic plots}
\kwd{Hill numbers}
\kwd{Power law}
\kwd{Rarefaction curve}
\kwd{Species-time relationship}
\kwd{Species richness}
\end{keyword}
\end{frontmatter}

\section{Introduction}
\label{s:Intro}

Estimating the diversity of classes in a population is a problem encountered in many fields. 
We may be interested in the diversity of words a person know from his/her writings \citep{efron1976}, the illegal immigrants from the apprehension records \citep{bohning2005}, the distinct attributes in a database 
\citep{haas1995,deolalikar2016}, or the distinct responses to a crowdsourcing query \citep{trushkowsky2012}. Among different applications, species abundance is the one that receives most attention. For this reason, it is chosen as the theme of this paper with the understanding that the proposed framework and methods are applicable in other applications as well. 

Understanding the species abundance in an ecological community has
long been an important task for ecologists. Such knowledge
is paramount in conservation planning and biodiversity management
\citep{matthews2015species}. An exhaustive species inventory is too labor
and resource intensive to be practical. Information about species
abundance can thus be acquired mainly through a survey.

Let $N=(N_{0},N_{1},...)$ where $N_{k}$ is the number of
species in a community that are represented exactly $k$ times in a survey. We do not observe the whole $N$,
but $\tilde N=(N_{1},N_{2},...)$ which is the zero-truncated
$N$. In other words, we do not know how many species are
not seen in the survey. We call the vector $\tilde N$,
the {\em frequency of frequencies} (FoF) \citep{good1953}.
 
A plethora of species abundance models have been proposed for $\tilde N$. Comprehensive review
of the field can be found in 
\citet{bunge1993} and \citet{matthews2014fitting}. A typical assumption in purely statistical models is  
\begin{equation}
\tilde N \mid N_+ \sim \mathrm{Multinomial}(N_+,p), \label{eq:multM}
\end{equation}
where $N_+ = \sum_{k=1}^\infty N_k$ is the total number of recorded species, and $p=(p_1,p_2,\ldots)$ is a probability vector, such that $p_k$ is the probability for a randomly selected recorded species to be observed $k$ times in the survey for $k=1,2, \ldots$. We call this vector $p$, the {\em species abundance distribution} (SAD). The great significance of (\ref{eq:multM}) relies on three assumptions: (A1) the species names are noninformative; (A2) the observed data for different recorded species are independent and identically distributed (iid), and (A3) for each recorded species, its observed frequency contains all useful information. 

Species accumulation curve (SAC) is another popular tool in the analysis of species abundance data. The survey is viewed as a data-collection process in which more and more sampling effort is devoted. The individual-based SAC is the number of recorded species expressed as a function of the amount of sampling effort. 

Despite the different emphases of SAD and SAC, the two approaches have an overlapped target: Estimating $D$, the total number of species in the community. For SAD, it means estimating $N_0$, the number of unseen species as $D=N_0+N_+$. For SAC, $D$ is the total number of seen species when the sampling effort is unlimited. 

As SAD depends largely on the sampling effort, it is necessary to include sampling effort explicitly in the model in order to make comparison of different SADs possible. This addition  establishes a link between SAD and SAC. Sampling effort can be of continuous type, such as the
area of land or the volume of water sampled, or the duration of
the survey. Discrete type sampling effort can be the sample size. To emphasize the sampling effort considered, the SAC is called {\em species-time curve}, {\em species-area curve}, or {\em species-sample-size curve} when the sampling effort is time, area, or sample size respectively. Species-area curve has been studied extensively in the literature.
Review of it can be found in \citet{tjorve2003shapes,tjorve2009shapes,dengler2009function} and \citet{williams2009}. In this paper, we use time as the measure of sampling effort. We consider the vector $N$ to be a function of the time $t$, denoted as $N(t)$. Notations $N_k$, $\tilde N$,
$N_+$, $p$ and $p_k$ are likewise denoted as $N_k(t)$, $\tilde N(t)$, $N_+(t)$,
$p(t)$ and $p_k(t)$ respectively. The (empirical) SAC is $N_+(t)$. Throughout the paper, we assume without loss of generality that the survey starts at time $0$ and ends at time $t_0 > 0$. 
Because of the great similarity of the species-time relationship and species-area
relationship \citep{preston1960time,magurran2007species}, time and area are treated as if they are interchangeable in this paper. When SAC is a species-area curve, we refer to the Type I species-area curve \citep{scheiner2003six}, where the areas sampled are nested (smaller areas are included in larger areas), resulting in a curve that is always nondecreasing.

The study of the relation between the (empirical) SAD and the (empirical) SAC started early since the introduction of rarefaction curve (\citep{arrhenius1921species,sanders1968}) which describes how we interpolate SAC using the observed FoF. Let $n_{k}(t_{0})$ be the observed $N_{k}(t_{0})$ for $k=1,2,\ldots$, $n_+(t_0)=\sum_{k=1}^\infty n_k(t_0)$, and $\tilde n(t_0) =\{n_k(t_0)\}_{k\geq 1}$. The rarefaction curve for species-time relationship is
\begin{equation}
\hat N_+(t) = \sum_{k=1}^\infty n_k(t_0)\left(1 - \left(1-\frac{t}{t_0} \right)^k \right),~~~ (0 \leq t \leq t_0). \label{eq:RF}
\end{equation}
The rationale of (\ref{eq:RF}) is that $1-(1-t/t_0)^k$ is the probability for a species with frequency $k$ in time interval $[0,t_0]$ to be observed before time $t$ if the appearance times of it in $[0,t_0]$ are iid $U[0,t_0]$ distributed. Equivalent formula for species-area curve appears earlier in \citet{arrhenius1921species}. \citet{good1956number} proposed an extrapolation formula for species-sample-size curve which when expressed as species-time curve is 
\begin{equation}
\hat N_+(t) = n_+(t_0) + \sum_{k=1}^\infty (-1)^{k+1}n_k(t_0)\left( \frac{t}{t_0} -1\right)^k,~~~(t > t_0). \label{eq:GTE}
\end{equation}
Equation (\ref{eq:GTE}) is (\ref{eq:RF}) with a change of domain. Define the empirical SAD, $\hat p_k(t) = n_k(t)/n_+(t)$. Equation (\ref{eq:GTE}) becomes
\begin{equation}
\hat N_+(t) = n_+(t_0)\left[1 - \sum_{k=1}^\infty \hat p_k(t_0)\left(1- \frac{t}{t_0}\right)^k \right] = n_+(t_0) \left[1- \hat h_{t_0}\left(1 -\frac{t}{t_0} \right) \right],  \label{eq:rel}
\end{equation}
where $\hat h_t(s)$ is an estimator of $h_t(s)$, the probability generating function  of the SAD at time $t$. Equation (\ref{eq:rel}) delineates a simple and yet convincing formula to find SAC from SAD and vice versa. The aim of this paper is to propose and study a statistical framework under which model in (\ref{eq:multM}) and the relation in (\ref{eq:rel}) hold when $\hat  N_+(t)$, $n_+(t_0)$ and $\hat h_{t_0}(s)$ are replaced by $E(N_+(t))$, $E(N_+(t_0))$ and $h_{t_0}(s)$ respectively. Our framework assumes (A1), (A2) and that the appearance times of each species follow a Poisson process which is sufficient for (A3) and the validity of (\ref{eq:RF}). 

Finite $D$ is commonplace in SAD approach although no consensus has been reached.  
Empirically, log-series distribution, an SAD that assumes infinite $D$, is one of the most successful models. Many SACs do not have asymptote. It is common that the observed number of rare species is large, and shows no sign to decrease. In Bayesian nonparametric approach in genomic diversity study, Poisson-Dirichlet process which assumes infinite $D$ is used (see for example \citet{lijoi2007}). In reality, the existence of transient species (species which are observed erratically and infrequently \citep{novotny2000}), and the error in the species identification process (a well-known example is the missequencing in pyrosequencing of DNA (see for example \citet{dickie2010insidious})) are continual sources of rare species making the species number larger than expected. In our framework, $D$ is random and can be finite or infinite with probability one. We use species richness to refer to $E(D)$ when $D$ is random, and $D$ when $D$ is deterministic. 

A special feature of our framework is that species can have zero detection probability. We call such species, zero-rate species, and all other species, positive-rate species. Zero-rate species can either be seen only once or unseen in a survey. If there are finite number of zero-rate species, the probability of observing any of them is zero. Therefore, if zero-rate species is observed in a survey, almost surely there are infinite number of them. We interpret observed individuals of positive-rate species as outcomes of a discrete distribution, where individuals of the same species can appear any nonnegative number of times in a survey. On the other hand, individuals of zero-rate species are outcomes of a continuous distribution, and no two such individuals belong to the same species. In our framework, the distribution is allowed to be a mixture of the two. Suppose we want to estimate the population of a town through recording each person we meet on street. Then tourists from distant countries can be viewed as ``zero-rate species''.
In the first example in Section \ref{s:swine}, it is suspected that the sequencing error in pyrosequencing of DNA may be a source for zero-rate species.

Poisson distribution is a main component in our framework. Though Poisson distribution is common in existing SAD models, time $t$ scarcely plays a role. In mixed Poisson model (see for example \citet{fisher1943relation,bulmer1974}), the observed frequencies of the species are independent and each follows a Poisson distribution with its own rate, say $\lambda_i$ for species $i$. The value $D$ is a fixed unknown finite value and $\{\lambda_i\}$ are iid sample from a mixing distribution. 
Another related model was proposed in \citet{zhou2017frequency}. The paper focuses mainly on finite $D$. Neither time nor zero-rate species are included in the model.

The outline of this paper is as follows. We propose in Section~\ref{s:MPPP} a new model for the sampling process, called the mixed Poisson partition process (MPPP). 
We emphasize on the parametric approach where additional assumptions are made on top of the framework so that the process depends only on a few parameters. Once the parameters are estimated, we can make inference on different characteristics of the population, say the SAD at any fixed time, the Hill numbers, or the expected future data. We study $\tilde N(t)$ in Section~\ref{s:FoF}. In Section~\ref{s:ESAC}, we consider the expected species accumulation curve (ESAC). We prove the one-to-one correspondence among (i) an MPPP, (ii) an ESAC which is a Bernstein function that passes through the origin, and (iii) the expected number of recorded species and the SAD at time $t_0$ such that the first derivative of its probability generating function is absolutely monotone in $(-\infty,1)$.
In Section~\ref{s:LDRF}, we introduce $\mathrm{LDR}_1$, a parametric family of ESAC. The SAD of $\mathrm{LDR}_1$ is the Engen's extended negative binomial distribution. This family has an attractive property that the ratio of the first and the second derivatives of the ESAC is a linear function of $t$. We extend $\mathrm{LDR}_1$ to $\mathrm{LDR}_2$ which allows zero-rate species. In Section \ref{s:Diag}, a D1/D2 plot for $\mathrm{LDR}_1$ and some diagnostic plots for specified $\mathrm{LDR}_1$ distributions are proposed. Extrapolation of the curve in D1/D2 plot is considered in Section \ref{s:Local}. Estimator of species richness is suggested basing on a modified first-order extrapolation. In Section~\ref{s:extend}, $\mathrm{LDR}_1$ is generalized so that the derivative ratio is a rational function instead of a linear function of $t$. Four real data are analyzed in Section~\ref{s:swine} to demonstrate the applications of the proposed models and the suggested plots.
In Section~\ref{s:SD}, we propose and study a design where only a few leading appearance times of each species are recorded. 
In Section \ref{s:curve2}, we consider the maximum likelihood approach on the empirical SAC. 
We give a discussion in Section \ref{s:discuss}. 

\section{Mixed Poisson Partition Process}
\label{s:MPPP}

Poisson process is the backbone of the framework. A Poisson process is a point process characterized by an intensity measure over the $n$-dimensional space $\mathbb{R}^n$ (we usually have $n=1$ in this paper). The intensity measure which we denote as $\omega$ delineates how many points are present on average in different parts of $\mathbb{R}^n$. More precisely, the number of points in a set $ A \subseteq \mathbb{R}^n$ follows the $\mathrm{Poisson}(\omega(A))$ distribution. Furthermore, for any finite collection of disjoint subsets $A_1, \ldots, A_k \subseteq \mathbb{R}^n$, $S_1, \ldots, S_k$ are mutually independent, where $S_i$ is the number of points in $A_i$. If $\omega(dx)=f(x) dx$, we call $f(x)$ the intensity function. A simple example is the homogeneous Poisson process where $\omega(dx)=\lambda dx$ for a constant $\lambda$. The parameter $\lambda$ is called the {\em rate} of the process.

If $\omega$ is finite (i.e., $\int \omega(dx) < \infty$), simulation of a Poisson process can be performed in two steps: (i) simulate the total number of points $W \sim \mathrm{Poisson}(\int \omega(dx))$, and (ii) simulate $X_1, \ldots, X_W$ iid from the probability measure $\omega /\int \omega (dx)$. Nevertheless, if $\omega$ is infinite, then the number of points is infinite, and it is impossible to simulate all the points in the process. In this case, we can only simulate a selected finite subset of points of the process using the {\em thinning property} of Poisson processes \citep{kingman1992poisson}. For each point $x$ in the Poisson process, let $\alpha(x) \in [0,1]$ be the probability for the point $x$ to be selected, and $1-\alpha(x)$ be the probability for it to be discarded. Then the selected points form a Poisson process with intensity measure $\alpha \omega$, where $\alpha \omega(A) = \int_A \alpha(x) \omega(dx)$. We can simulate the selected points when the measure $\alpha \omega$ is finite using the aforementioned method. For example, if we are interested only in the points that lie in a bounded region $A \subset \mathbb{R}^n$ of a homogeneous Poisson process with rate $\lambda$, then $\alpha(x) = \mathbf{1}\{x \in A\}$ where $\mathbf{1}\{.\}$ is the indicator function. The number of selected points follows $\mathrm{Poisson}(\lambda \int_A dx)$ distribution, and the selected points are iid uniformly distributed in $A$.

We model the observations in a species abundance survey by a stochastic process where rates of the species follows a Poisson process. 

\vspace{0.3cm}
\noindent
{\bf Definition:} (Mixed Poisson partition process)
\label{def:mixed_poisson} A mixed
Poisson partition process (MPPP), $G$, is characterized by a nonzero species intensity
measure $\nu$, which is a measure over $\mathbb{R}_{\ge0}$, the set of all nonnegative real numbers, satisfying
\begin{equation}
\int_{0}^{\infty}\min\{1,\lambda^{-1}\}\nu(d\lambda)<\infty.\label{eq:nu_finite}
\end{equation}
The definition of an MPPP consists of three
steps: 
\begin{enumerate}
\item[1.] (Generation of positive rates of species) Given $\nu$, define $\tilde{\nu}$
to be a measure over $\mathbb{R}_{>0}$ ($\mathbb{R}_{>0}$ is the
set of positive real numbers) by $\tilde{\nu}(d\lambda)=\nu(d\lambda)/\lambda$,
(i.e., $d\tilde{\nu}/d\nu=1/\lambda$ for $\lambda>0$). 
Let $\lambda_{1},\lambda_{2},\ldots$ (a finite or countably
infinite sequence) be a realization of a Poisson process with intensity measure
$\tilde{\nu}$. 
\item[2.] (Generation of individuals of positive-rate species) For each $\lambda_i$ in Step 1, we
generate a realization $\eta_i$ (independently across $i$) of a Poisson process with rate
$\lambda_{i}$. The realization $\eta_i$ represents the arrival times of a species with rate $\lambda_i$.
\item[3.] (Generation of individuals of zero-rate species) We generate a realization $\eta_0$ of a Poisson process with rate $\nu(\{0\})$, independent
of $\eta_{1},\eta_{2},\ldots$. This represents the times of appearance for all the zero-rate species.
\end{enumerate}
Finally, we take $G=\{\eta_{1},\eta_{2},\ldots\}\cup\eta_{0}$. For any $i\ge 1$, all points in $\eta_{i}$ are arrival times of the same
species, whereas each point in $\eta_{0}$ is from a different zero-rate species
(we use a slight abuse of notation to treat each point in $\eta_{0}$
as a point process with only one point).

\vspace{0.3cm}
Measures $\nu$ and $\tilde{\nu}$ may be finite or infinite. Let $\Lambda=\int \nu(d \lambda)$. As the
expected number of individuals seen in time interval $[0,t]$ is equal
to $t \Lambda$ (see \eqref{eq:finite-nu} in Section~\ref{s:FoF}), we can interpret $\Lambda$ as the expected total rate. When $\nu$ is finite (i.e., $\Lambda < \infty$), conditional on the event that there is an individual observed exactly at time $t$, the distribution of the rate of the species of that individual is $\nu/\Lambda$ (see Proposition A in Appendix \ref{AppFix} for a proof). Therefore, we can regard $\nu$ as the intensity measure of $\lambda$ of the observed individuals. If $\tilde{\nu}$ is finite, the expected number
of positive-rate species in the community is
finite. From Step 3 in the definition, if $\nu(\{0\})>0$, there are
infinite number of zero-rate species and they arrive at a constant
rate. With probability one, $D$ is finite if and only if $\nu(\{0\})=0$
and $\tilde{\nu}(\mathbb{R}_{>0})<\infty$. In such case, $D$ follows a 
$\mathrm{Poisson}(\tilde{\nu}(\mathbb{R}_{>0}))$ distribution.
Measure $\tilde \nu$ specifies the distribution of the rates of positive-rate species. More precisely, $\tilde \nu([\lambda_0, \lambda_1])$ with $\lambda_0 > 0$ is the average number of species with rate in $[\lambda_0, \lambda_1]$. Measure $\nu$ specifies the distribution of the rates of the species of the individuals including those of the zero-rate species. More precisely, $ \nu([\lambda_0, \lambda_1])$ is the average number of individuals (per unit time) belonging to species whose rates lie in $[\lambda_0, \lambda_1]$. Condition \eqref{eq:nu_finite} is essential because it is equivalent to the finiteness of the ESAC (i.e., $E(N_+(t)) < \infty$ for any finite nonnegative $t$). A proof of it is given in Appendix \ref{AppB}. Figure~\ref{f:figure1} illustrates the generation of the MPPP. 

When $\tilde \nu$ is infinite, we cannot simulate all $\lambda_i$'s in Step 1 in practice. We can use the following method to simulate a realization of the process in interval $[0,t_0]$. The probability for a species with rate $\lambda$ to be recorded in $[0,t_0]$ is $1- \exp(-\lambda t_0)$. Applying the thinning property, the intensity of the recorded positive-rate species is $(1-\exp(-\lambda t_0)) \tilde \nu$ which is always finite. To include also the recorded zero-rate species, the intensity is $(1-\exp(-\lambda t_0)) \lambda^{-1} \nu$ (we take $(1-\exp(-\lambda t_0))\lambda^{-1} = t_0$ when $\lambda = 0$). After generating $\lambda_1, \lambda_2, \ldots$ according to a Poisson process with this intensity measure, we simulate for each $i$, a realization $\eta_i$ (independently across $i$) of a Poisson process with rate $\lambda_i$, conditional on the event that $\eta_i$ has at least one point in $[0,t_0]$ (if $\lambda_i=0$, then $\eta_i$ contains one uniformly distributed point in $[0,t_0]$). 
An alternative equivalent definition that unifies the generation of individuals of zero-rate species and positive-rate species is given in Appendix \ref{AppA}. 

\begin{figure}
\begin{center}
\includegraphics[width=5in]{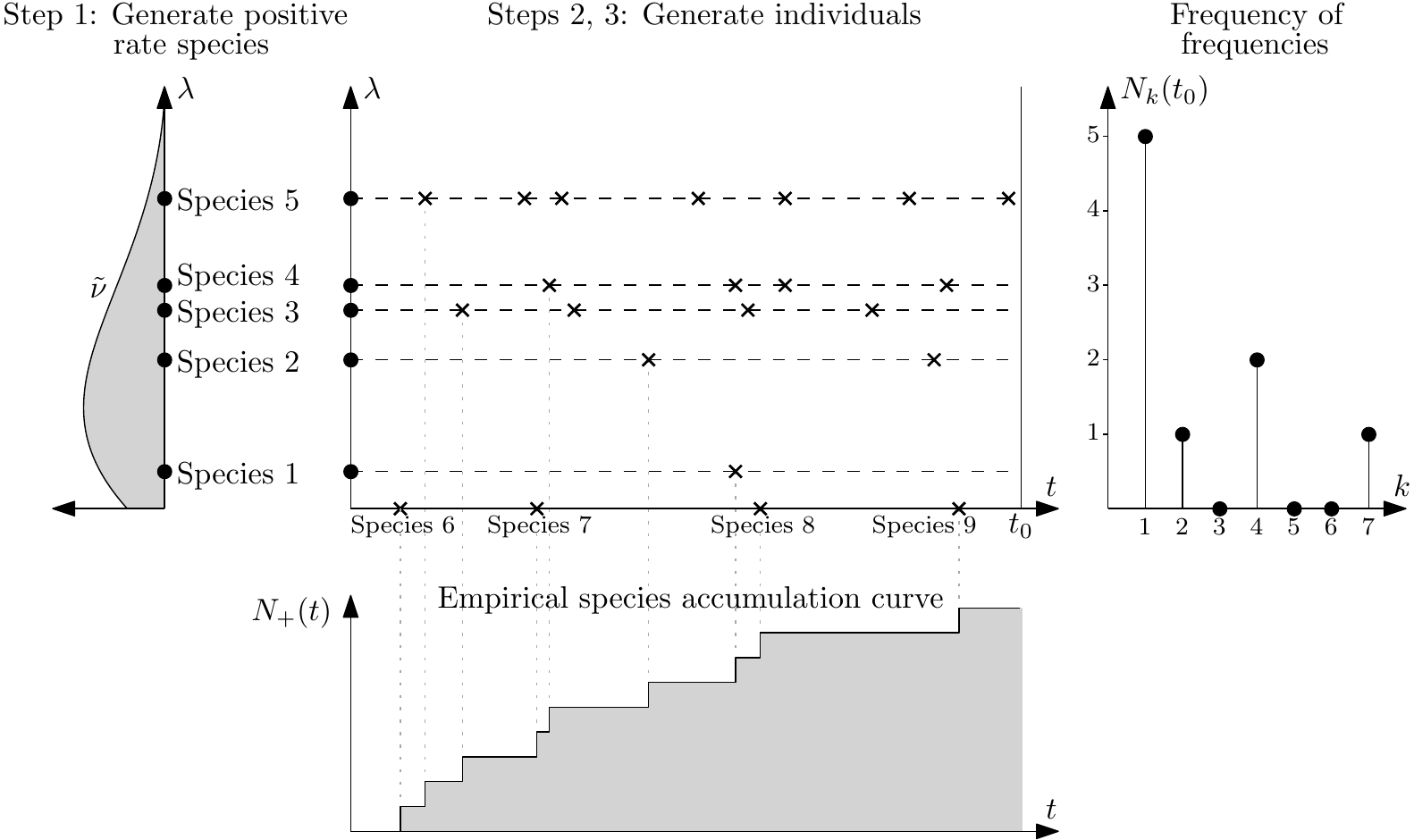}
\end{center}
\caption{An illustration of the MPPP. In step 1,
we generate the rates $\lambda_{i}$ of the positive-rate species
according to a Poisson process with intensity measure $\tilde{\nu}$.
If $\tilde{\nu}(\mathbb{R}_{>0})<\infty$, then this is equivalent to first
generating $n_{pos}$, the number of positive-rate species according to $\mathrm{Poisson}(\tilde{\nu}(\mathbb{R}_{>0}))$,
and then generating a random sample $\lambda_{1}, , \ldots, \lambda_{n_{pos}}$ of size $n_{pos}$ from the probability
measure $\tilde{\nu}/\tilde{\nu}(\mathbb{R}_{>0})$. In step 2, we generate
the individuals of species $i$ according to a Poisson process with
rate $\lambda_{i}$ for $i=1, \ldots, n_{pos}$. In step 3, we generate the individuals of zero-rate
species (species 6 to 9 in the figure) according to a Poisson process
with rate $\nu(\{0\})$. }
\label{f:figure1}
\end{figure}

A special feature of the framework is that $D$ is random and can be infinite with probability one. This change necessitates modification of biodiversity measures, among which Hill numbers are popular. When $D$ is deterministic, Hill number of order $q$ \citep{hill1973} for $q \geq 0$ and $q \neq 1$ is $^{q}\!D=\Big(\sum_{i}(p_{\mathrm{sp}}(i))^{q}\Big)^{1/(1-q)}$
where $p_{\mathrm{sp}}(i)$ is the relative abundance
of species $i$ in an assemblage (the number of individuals of species
$i$ divided by the total population).
When $q=1$, $^1\!D$ is defined as $\lim_{q \to 1}\, ^q\!D = \exp(-\sum_i p_{\mathrm{sp}}(i) \log(p_{\mathrm{sp}}(i)))$. Hill numbers are interpreted as the effective number of species. 
The order $q$ determines the sensitivity of the measure to species relative abundances. When $q=0$, all species regardless its abundance are treated equally. Larger $q$ imposes more weight to abundant species. Hill numbers encompass three important diversity measures as its special cases. They are the species richness ($q=0$), the exponential of Shannon entropy ($q=1$), and the inverse of Simpson index ($q=2$). A notable property of Hill numbers is that they obey the replication principle: The diversity of an assemblage formed by pooling $m$  equally abundant and equally large assemblages with no species in common is $m$ times the diversity of a single assemblage. 

Under our framework, each species corresponds to a Poisson process, and the relative abundance of a species is its rate divided by the expected total rate, $\Lambda$.  
Reasonable modifications to the definition of Hill numbers are to replace $p_{\mathrm{sp}}(i)$ with $\lambda/\Lambda$,
where $\lambda$ is the rate of a species, and to replace summation
over species with integration with respect to the measure $\lambda^{-1}\nu$ so that the integral of $\lambda/\Lambda$ is one. It works well when $\Lambda$ is finite, but fails when $\Lambda$ is infinite. To fix the problem, we need a meaningful surrogate rate which fulfills two requirements: (i) the expected total rate is finite, and (ii) it approaches the true rate $\lambda$ as a limit. The first appearance time of a species with rate $\lambda$ follows the exponential distribution with density function  $\lambda\exp(-\lambda t)$, which can be regarded as the instantaneous
rate of its first appearance at time $t$. This rate approaches $\lambda$ when $t$ decreases to zero. The expected total 
instantaneous rate of first appearances over all species at time $t$
is $\Lambda_{t}=\int\exp(-\lambda t)\nu(d\lambda)$ ($=E(N_1(t))/t$ from (\ref{eq:Enk})) which is always finite for positive $t$. 
Replacing $p_{\mathrm{sp}}(i)$
with $\lambda\exp(-\lambda t)/\Lambda_{t}$ and taking $t\to 0$, we
define, the Hill numbers, $^q\! D_\nu$ for our framework as
\[
^{q}\! D_\nu= \begin{cases} \lim_{t\to 0}\bigg(\Lambda_{t}^{-q}\int\lambda^{q-1}\exp(-\lambda qt)\nu(d\lambda)\bigg)^{1/(1-q)} & (q \geq 0, q \neq 1) \\ 
\lim_{t\to 0}\Lambda_{t}\exp\bigg(-\frac{1}{\Lambda_{t}}\int \left(\log\lambda-\lambda t\right) \exp(-\lambda t)\nu(d\lambda)\bigg) & (q=1). \end{cases}
\]
When $\nu(\{0\}) > 0$ and $0 \leq q \leq 1$, $^q\!D_\nu = \infty$. When $\Lambda$ is finite, 
\[
^{q}\! D_\nu= \begin{cases} \bigg(\Lambda^{-q}\int\lambda^{q-1}\nu(d\lambda)\bigg)^{1/(1-q)} & (q \geq 0, q \neq 1) \\
\Lambda \exp \bigg(-\frac{1}{\Lambda}\int \log(\lambda) \nu(d\lambda)\bigg) & (q=1).
\end{cases}
\]
Diversity $^q\! D_\nu$ is non-increasing with respect to $q$. Unlike the classical Hill numbers, $^q\!D_\nu$ can be less than one. For instance, from (\ref{eq:finiteD}), $^0\! D_\nu = E(D)$ which can be any positive value. It can be proved that for any positive constant $\gamma$, $^q\! D_\nu = \gamma (^q\! D_{\nu/\gamma})$, an analogue of the replication principle for Hill numbers.

\section{Frequency of Frequencies}
\label{s:FoF}

A sufficient statistic for a realization $G$ of an MPPP in time interval $[0,t_0]$ is $\tilde N(t_0)$. Consider a time interval $[0,t]$. For each $\lambda$ in Step 2 of the definition, it contributes one to one of the counts $N_k(t)$, where $k$ follows a Poisson distribution, $\mathrm{Poisson}(\lambda t)$. By the splitting property of Poisson processes \citep{kingman1992poisson}, their total contribution to $N_k(t)$ follows $\mathrm{Poisson}(\int(\lambda t)^k(k!)^{-1}e^{-\lambda t} \tilde \nu(d \lambda))$ distribution independent across $k$. The zero-rate species only increase $N_1(t)$ by a $\mathrm{Poisson}(\nu(\{0\})t)$ random variate. Therefore, for $k \geq 1$, 
\begin{eqnarray*}
N_k(t) & \sim & \mathrm{Poisson} \left (\int \frac{(\lambda t)^k \exp(-\lambda t)}{k!} \tilde \nu(d \lambda) + \mathbf{1}\{k=1\}\nu(\{0\})t \right) \\
& = & \mathrm{Poisson} \left( \int \frac{\lambda^{k-1} t^k \exp(-\lambda t)}{k!} \nu(d \lambda) \right),
\end{eqnarray*}
(we use the convention that $0^0=1$).
Equations (\ref{eq:En+}) and (\ref{eq:finiteD}) follows from $(\ref{eq:Enk})$ for $k \geq 1$. The correctness of (\ref{eq:Enk}) when $k=0$ follows from (\ref{eq:En+}), (\ref{eq:finiteD}) and the fact $E(N_0(t)) = E(D) - E(N_+(t))$.  
\begin{eqnarray}
E(N_{k}(t)) & = & \int\frac{\lambda^{k-1}t^{k} \exp(-\lambda t)}{k!}\nu(d\lambda),~~~(k \geq 0) \label{eq:Enk} \\
E(N_{+}(t)) & = & \sum_{k=1}^\infty E(N_k(t))= 
\int\frac{1-\exp(-\lambda t)}{\lambda}\nu(d\lambda), \label{eq:En+} \\
E(D) & = & \lim_{t\to\infty}E(N_{+}(t)) = \int \lambda^{-1} \nu(d \lambda). \label{eq:finiteD}
\end{eqnarray}
Again for (\ref{eq:En+}), we set $(1-\exp(-\lambda t))/\lambda = t$ when $\lambda = 0$. When $\nu(\{0\})>0$, $E(D)$ is infinite. Since all elements in $\{N_k(t)\}_{k\geq 1}$ are independent and follow Poisson distribution, variable $N_+(t)$ is Poisson distributed, and so do $D$ and $N_0(t)$ when their expected values are finite.

Write $S(t) = \sum_{k=1}^\infty k N_k(t)$ which is the number of individuals observed before time $t$. 
\begin{equation}
E(S(t)) = \int \sum_{k=1}^\infty \frac{\lambda^{k-1}t^k \exp(-\lambda t)}{(k-1)!} \nu (d \lambda) = t\int\nu(d\lambda) = t \Lambda. \label{eq:finite-nu}
\end{equation}

From (\ref{eq:Enk}), for $k \geq 0$, 
\[
\frac{(k+1)E(N_{k+1}(t_0))}{t_0} = \int \lambda \frac{(\lambda t_0)^{k} \exp(-\lambda t)}{k! \lambda}\nu(d\lambda).
\]
By the thinning property of Poisson processes,
\[
\frac{(\lambda t_0)^{k} \exp(-\lambda t)}{k! \lambda}\nu(d\lambda)
\]
is the intensity measure of the rate for the species represented $k$ times in $[0,t_0]$. We can interpret $(k+1)E(N_{k+1}(t_0))/t_0$ as the expected total rate for all species represented $k$ times in $[0,t_0]$. As pointed out in Section \ref{s:MPPP} and from (\ref{eq:finite-nu}), $\Lambda = E(S(t_0))/t_0$ is the expected total rate for all species. 
Conditional on the event that we will observe an individual at a given future time $t_1 > t_0$, the probability that this individual belongs to a species represented $k$ times in $[0,t_0]$ is equal to 
\[
\frac{(k+1)E(N_{k+1}(t_0))/t_0}{E(S(t_0))/t_0} = \frac{(k+1)E(N_{k+1}(t_0))}{E(S(t_0))}
\]
(formal proof is given in Appendix \ref{AppL}) which corresponds to the renowned Good-Turing frequency estimator in \citet{good1953}. It is worth pointing out that the above equation holds for individual observed at a given future time rather than the next observed individual.

From the well-known relation between independent Poisson random variables and multinomial distribution, model (\ref{eq:multM}) is valid under the framework with  
\begin{equation}
p_k(t) = \frac{E(N_k(t))}{E(N_+(t))}=\frac{\int (k!)^{-1}\lambda^{k-1}t^{k}\exp(-\lambda t)\nu(d\lambda)}{\int \lambda^{-1}(1-\exp(-\lambda t)) \nu(d\lambda)}, ~~~~(k=1,2,\ldots). \label{eq:sad1}
\end{equation}
Formal proof is given in Appendix \ref{appprob}. If $E(D)$ is finite, $\lim_{t\to \infty} p_k(t)=0$ for any fixed $k$. 
The joint probability mass function of $\tilde N(t_0)$ is
\[
P(\tilde N(t_0)=\tilde n(t_0) \mid \nu)  = \exp\left(-E(N_{+}(t_{0}))\right)\prod_{k=1}^{\infty}\frac{(E(N_{k}(t_{0})))^{n_{k}(t_{0})}}{n_{k}(t_{0})!}.
\]
In terms of the expected FoF, the log-likelihood function is
\begin{equation}
\log(\mathcal{L}(\{E(N_k(t_0))\}_{k\geq 1} \mid \tilde n(t_0)))=-E(N_{+}(t_{0})) + \sum_{k=1}^{\infty}n_k(t_0) \log(E(N_{k}(t_{0}))).\label{eq:Lik}
\end{equation}
In terms of $p(t_0)$ and $E(N_+(t_0))$, it is
\begin{align*}
& \log(\mathcal{L}(p(t_0),E(N_+(t_0)) \mid \tilde n(t_0))) \\
& = -E(N_{+}(t_{0}))+n_+(t_0)\log(E(N_+(t_0)))+ \sum_{k=1}^{\infty} n_k(t_0) \log(p_{k}(t_{0})). 
\end{align*}
If the unknown vector $p(t_0)$ and the quantity $E(N_+(t_0))$ are unrelated, the above log-likelihood function implies that the maximum likelihood estimator (MLE) of $p(t_0)$ is the conditional maximum likelihood estimator (conditional on the observed $n_+(t_0)$) for the multinomial distribution in (\ref{eq:multM}). The MLE of $E(N_+(t_0))$ is $n_+(t_0)$. 

\section{Expected Species Accumulation Curve}
\label{s:ESAC}

Denote the expected (empirical) SAC (ESAC) as $\psi(t)=E(N_+(t))$.  
Condition (\ref{eq:nu_finite}) guarantees that $\psi(t)$ is finite for any finite $t$. For a real-valued function $g(t)$, let $g^{(m)}(t)$ stand for the $m$-order derivative of function
$g(t)$. Clearly $\psi(0)=0$. From \eqref{eq:En+}, 
\begin{equation}
\psi^{(k)}(t)=\int(-\lambda)^{k-1}\exp(-\lambda t)\nu(d\lambda),~~~(k = 1, 2, \ldots). \label{eq:esac_dk}
\end{equation}
Note that $\psi^{(1)}(0) = \int \nu(d \lambda) = \Lambda$. From \eqref{eq:Enk} and \eqref{eq:esac_dk}, 
\begin{equation}
\psi^{(k)}(t)=(-1)^{k+1}\frac{k!}{t^{k}}E(N_{k}(t)),~~~(k=1,2,\ldots). \label{eq:deriv-1}
\end{equation}
Analogous expression for \eqref{eq:deriv-1}
appears in \citet{beguinot2016extrapolation} as an approximate formula under the multinomial model for fixed total number of observed individuals
for the species-sample-size curve with
the derivative operator replaced by the difference operator. 

Before studying the link between ESAC and SAD, two mathematical terms are needed. A function $g(t)$ is a {\em Bernstein function} if it is a nonnegative real-valued function on $[0,\infty)$ such that $(-1)^{k+1}g^{(k)}(t) \geq 0$
for all positive integer $k$ \citep{schilling2012bernstein}. An infinitely differentiable function $f(s)$ on an interval $A$ is called {\em absolutely monotone} in $A$ if $f^{(k)}(s) \geq 0$ for $k=0, 1, \ldots$ and $s \in A$. The relation between absolutely monotone function and probability generating function is well known (see for example \citet{stroock2010}).  

\vspace{0.3cm}
\noindent
{\bf Proposition 1:}

\begin{em}
\begin{description}

\item[a.] A function $\psi(t)$ is the ESAC of an MPPP if and only if $\psi(t)$ is a Bernstein function that passes through the origin.

\item[b.] A function $h_t(s)$ is the probability generating function of $p(t)$ for a fixed positive $t$ of an MPPP if and only if $h_t(0)=0, h_t(1)=1$, and $h_t(s)$ is absolutely monotone in $(-\infty, 1)$. 

\item[c.] An MPPP is uniquely determined by its $\psi(t)$, or $(h_{t_0}(s), \psi(t_0))$ for the probability generating function $h_{t_0}(s)$ of $p(t_0)$ for a fixed positive $t_0$.

\item[d.] For any MPPP, and $t > 0$, we have 
\begin{equation}
h_t(s) = 1 - \frac{\psi((1-s)t)}{\psi(t)},~~~(s \in (-\infty, 1]). \label{eq:pgf1}
\end{equation}

\end{description}
\end{em}

\vspace{0.3cm} 
Proof of Proposition 1 is given in Appendix \ref{AppProp1}. Equation (\ref{eq:pgf1}) is the population version of (\ref{eq:rel}). 

It worths pointing out that  
\citet{boneh1998estimating} proved that $(-1)^{k+1}\psi^{(k)}(t) \geq 0$ for a different setting ($D$ independent Poisson processes with different rates) and called it ``infinite order alternating copositivity''. 

From \eqref{eq:Lik}, the log-likelihood function can be re-expressed as a function of $\psi(t)$.
\begin{equation}
\log(\mathcal{\mathcal{L}}(\psi \mid \tilde n(t_0)))=-\psi(t_{0}) + \sum_{k=1}^{\infty} n_k(t_0) \log(\mid \psi^{(k)}(t_{0}) \mid ). \label{eq:psiL}
\end{equation}

It can be shown that the Taylor expansion of $\psi(t)$ at $t_0$ converges to $\psi(t)$ when $0 \leq t < 2t_0$:
\begin{equation}
\psi(t) = E(N_{+}(t_{0}))+\sum_{k=1}^{\infty}(-1)^{k+1}E(N_{k}(t_{0}))\left(\frac{t}{t_{0}}-1\right)^{k}, ~~~(0 \leq t < 2t_0). \label{eq:psiv}
\end{equation}
It signifies that Good-Toulmin estimator in (\ref{eq:GTE}) performs satisfactorily in short-term extrapolation when $t_{0}<t < 2t_{0}$.
We deduce from (\ref{eq:psiv}) the following unbiased estimator for different order of derivative of $\psi(t)$ 
\begin{equation}
\hat{\psi}^{(j)}(t)=\frac{(-1)^{j+1}}{t_{0}^j}\sum_{k=j}^{\infty} \frac{k!n_k(t_0)}{(k-j)!} \left(1-\frac{t}{t_{0}}\right)^{k-j},~~~(j \geq 1, 0 < t < 2t_0). \label{eq:dpsi}
\end{equation}
We use (\ref{eq:dpsi}) only for $0 \leq t \leq t_0$ \footnote{We include $t=0$ here for completeness. Remember that $\psi^{(1)}(0) = \Lambda$ can be infinite.} because outside this interval, $\hat{\psi}^{(j)}(t)$ may not have the correct sign $(-1)^{j+1}$.

Estimator in (\ref{eq:dpsi}) is useful. For example, a concave downward curve when we plot $1/\hat{\psi}^{(1)}(t)$ for $t\in (0,t_{0}]$ \footnote{It corresponds to the diagnostic check for the log-series distribution in Table 1 in Section \ref{s:LDRF}.} is an indication that $E(D)=\infty$ because if we believe that there is a linear function
$b+ct$ with positive $b$ and $c$ such that $b+ct\geq1/\psi^{(1)}(t)$
for all $t\geq 0$, then
\[
E(D)=\int_{0}^{\infty}\psi^{(1)}(x)dx\geq\int_{0}^{\infty}(b+cx)^{-1}dx=\infty.
\]

From (\ref{eq:deriv-1}), a parallel result of Equation (\ref{eq:dpsi}) is the following  relation among expected FoFs 
\begin{equation}
E(N_j(t)) = \sum_{k=j}^\infty  E(N_k(t_0)) {k \choose j}\left(\frac{t}{t_0}\right)^j \left(1 - \frac{t}{t_0} \right)^{k-j},~~~(j \geq 1, 0 \leq t < 2t_0), \label{eq:iter}
\end{equation}
which can be proved using the law of iterated expectations when $0 \leq t \leq t_0$. Furthermore, for $0 \leq t \leq t_0$
\begin{equation}
E(N_j(t) \mid \tilde N(t_0)=\tilde n(t_0)) = \sum_{k=j}^\infty n_k(t_0) {k \choose j}\left(\frac{t}{t_0}\right)^j \left(1 - \frac{t}{t_0} \right)^{k-j},~~~(j \geq 1), \label{eq:nj}
\end{equation}

\section{Linear First Derivative Ratio Family} \label{s:LDRF}

MPPP is nonparametric in nature. It is defined by an intensity measure, an ESAC or an SAD. When parametric approach is preferred, we restrict our interest to a family of distributions in MPPP, say by putting constraints on the SAD. A way to portray an SAD is to delineate its probability ratio, $p_j(t)/p_{j+1}(t)$ for $j = 1, 2, \ldots$. It is equivalent to examine $-\psi^{(j)}(t)/\psi^{(j+1)}(t)= tp_j(t)/[(j+1)p_{j+1}(t)]$, which we call the {\em $j$th derivative ratio}. From (\ref{eq:esac_dk}) and the Cauchy-Schwarz inequality, for $j=1, 2, \ldots$ and $t \geq 0$, $\psi^{(j)}(t) \psi^{(j+2)}(t) \geq (\psi^{(j+1)}(t))^2$.
It deduces that $j$th derivative ratio is always a nonnegative nondecreasing function of $t$. Among all derivative ratios, the first derivative ratio is most important because $-\psi^{(1)}(t)/\psi^{(2)}(t) = - [d \log(\psi^{(1)}(t))/dt]^{-1}$ which relates to the logarithmic derivative of $\psi^{(1)}(t) = E(N_1(t))/t$, the expected total rate for unseen species at time $t$. The following proposition gives a sufficient condition for it.

\vspace{0.3cm}
\noindent
{\bf Proposition 2:} 
{\em A sufficient condition for a function $\xi(t)$ on $[0,\infty)$ to be the first derivative ratio for an MPPP is that (i) $\xi(t)$ is a Bernstein function, and (ii) $\xi(0) > 0$ or $\xi^{(1)}(0)>1$.}

\vspace{0.3cm}
\noindent
We prove Proposition 2 in Appendix \ref{AppC}. Hereafter we use $\xi(t)=-\psi^{(1)}(t)/\psi^{(2)}(t)$ to denote the first derivative ratio. 

The simplest nontrivial Bernstein function is the positive linear function on $[0,\infty)$. It suggests the following fundamental family of distributions in MPPP. 

\vspace{0.3cm}
\noindent
{\bf Definition: (Linear First Derivative Ratio Family)} An MPPP is said to belong to the {\em linear first derivative ratio family} (denoted as $\mathrm{LDR}_1$) if its first derivative ratio $\xi(t)$ takes a linear form $\xi(t) = b+ct$ for $b \geq 0$ and $c \geq 0$ such that $c > 1$ if $b=0$. 

\vspace{0.3cm} 
Family $\mathrm{LDR}_1$ encompasses some common SADs and ESACs. We list the characteristics of all models in $\mathrm{LDR}_1$ in Table \ref{ta}. When $b=0$, $c$ must be larger than 1 (see condition (ii) in Proposition 2), otherwise from the second row of Table \ref{ta}, $\lim_{b \downarrow 0} \psi(t)=\infty$ for finite $t$. In Table \ref{ta}, equality of transformed $\psi^{(1)}(t)$ for some models is presented. Such equalities can be used to produce diagnostic check for specified SAD in $\mathrm{LDR}_1$ when $\psi^{(1)}(t)$ is replaced by its estimator in (\ref{eq:dpsi}). 

\vspace{0.3cm}
{ \renewcommand{\arraystretch}{1.4}
\begin{table}
\begin{threeparttable}[b]
\caption{Models in $\mathrm{LDR}_1$ (in the expressions, ``$a$'' is a positive scale parameter)}
\begin{center}
\begin{tabular}{ll}
\hline
$c=0$ & SAD (Zero-truncated Poisson distribution)\tnote{1}~~: \\
($ \Rightarrow b > 0$) & ~~~$p_k(t)=(t/b)^k \exp(-t/b)/[k!(1-\exp(-t/b))]$ \\
& ESAC (Negative exponential law): \\ 
& ~~~$\psi(t)=ab\exp(1/b)(1-\exp(-t/b))$ \\
& Diagnostic check:  $\log(\psi^{(1)}(t))=1/b+\log(a)-t/b$ \\
\hline
$ c \neq 0, 1$ & SAD\tnote{2}~~: $p_k(t) = 
\frac{(c-1)c^{k-1}t^k \Gamma(1/c+k-1) (b+ct)^{1-1/c-k}}{k!\Gamma(1/c)[(b+ct)^{1-1/c}-b^{1-1/c}]}$ \\
$b > 0$ & ESAC:  $\psi(t)=\frac{a(b+c)}{c-1}\left(\left(\frac{b+ct}{b+c}\right)^{1-1/c}-\left(\frac{b}{b+c}\right)^{1-1/c}\right)$ \\
\hline
$c=1/2$ & SAD (Geometric distribution)\tnote{3}~~: $p_k(t) = (2b/t) [t/(2b+t)]^k$ \\
($ \Rightarrow b > 0$) & ESAC (Hyperbola law)\tnote{4}~~:  $\psi(t)=a(2b+1)^{2}t/(2b(t+2b))$ \\
& Diagnostic check:  $(\psi^{(1)}(t))^{-1/2}=(t+2b)/[a^{1/2}(2b+1)]$ \\
\hline
$c=1$ & SAD (Log-series distribution): $p_k(t)=[t/(t+b)]^k/(k \log(1+t/b))$ \\
($ \Rightarrow b > 0$) & ESAC (Kobayashi's logarithm law)\tnote{5}~~: $\psi(t)=a(b+1)\log(1+t/b)$ \\
& Diagnostic check: $1/\psi^{(1)}(t)=(b+t)/[a(b+1)]$ \\
\hline
$b=0$  & SAD \tnote{6}~~: When $1 < c < \infty$, $p_k(t) = (c-1)\Gamma(1/c+k-1)/[k!c\Gamma(1/c)]$ \\
($ \Rightarrow c > 1$ & ~~~~~When $c=\infty$, $p_1(t)=1$, $p_k(t)=0$ for $k > 1$. \\
or $c=\infty$ ) & ESAC (Power law): When $1 < c < \infty$, $\psi(t)=act^{1-1/c}/(c-1)$ \\
& ~~~~~When $c=\infty$, $\psi(t)=at$ \\
& Diagnostic check:  $\log(\psi^{(1)}(t))=\log(a)-\log(t)/c$ \\
\hline
\end{tabular}
\begin{tablenotes}
\item [1] It is the simplest MPPP with all species having the same rate $1/b$.

\item [2] When $0 < c < 1$ ($\Rightarrow b > 0$), it is the zero-truncated negative binomial distribution.

\item[3] It is a special case of the zero-truncated negative binomial distribution.

\item[4] Also known as Michaelis-Menten equation and Monod model.

\item[5] \citet{kobayashi1975species} (see also \citet{fisher1943relation} and \citet{may1975patterns})

\item[6] This distribution appears in \citet{zhou2017frequency}.
\end{tablenotes}
\label{ta}
\end{center}
\end{threeparttable}
\end{table}
}

$\mathrm{LDR}_1$ has three parameters $a$, $b$ and $c$. Parameters $b$ and $c$ determine the SAD, and $a = \psi^{(1)}(1)$ is a scale parameter. As pointed out at the end of Section \ref{s:FoF}, the MLE of $b$ and $c$ is equivalent to the conditional MLE (conditional on the observed $n_+(t_0)$) of the multinomial model in (\ref{eq:multM}). The MLE of $a$ is chosen to make $\hat E(N_+(t_0))$ equal to $n_+(t_0)$.

From Table \ref{ta}, we can find $E(N_k(t))$ and $\psi^{(k)}(t)$ using the relations $E(N_k(t)) = p_k(t) \psi(t)$, and 
\[
\psi^{(k)}(t) = (-1)^{k+1} \frac{k!}{t^k} p_k(t) \psi(t).
\]
It can be shown that for $\mathrm{LDR}_1$, $\nu(\{0\})=0$, and  
\begin{equation}
\tilde \nu(d\lambda)=\begin{cases}
\frac{a((b+c)/c)^{1/c}\lambda^{1/c-2}}{\Gamma(1/c)}\exp(-b\lambda/c)d\lambda & (c>0),\\
ab\exp(1/b)\delta_{1/b}(d\lambda) & (c=0), \label{eq:nu2}
\end{cases}
\end{equation}
where $\delta_{1/b}(A)=\mathbf{1}\{1/b\in A\}$ is the Dirac measure. The intensity measure $\tilde \nu$ takes the form as a gamma distribution with extended shape
parameter $1/c-1$ for nonnegative $c$. Therefore, $p(t)$ is 
the Engen's extended negative binomial distribution \citep{engen1974species} with support $\{1,2,\ldots\}$. The parameter $c$ determines the shape of the gamma distribution.

The Hill number of order $q$ for $\mathrm{LDR}_1$ is
\[
^q\!D_\nu = \begin{cases} ab\exp(1/b) & (c=0) \\ a ((b+c)/b)^{1/c} (b/c) (\Gamma(\frac{1}{c}+q-1)/\Gamma(\frac{1}{c}))^{1/(1-q)} & (c > 0, q > 1-\frac{1}{c}, q \neq 1) \\ a ((b+c)/b)^{1/c} (b/c) \exp(-\Psi(\frac{1}{c})) & (c >0, q = 1) \\ \infty & (c>0, q \leq 1-\frac{1}{c}), \end{cases} 
\]
where $\Psi(x)$ is the digamma function. 
$E(D)$ can be found either as $^0\!D_\nu$ or $\lim_{t \to \infty} \psi(t)$. Species richness, $E(D)=\,^0\!D_\nu=\infty$ if and only if $c \geq 1$. In this case, $E(N_k(t))$ is increasing in $t$ for any fixed $k$. When $E(D)<\infty$, $E(N_k(t))$ is unimodal with respect to $t$. 

An ESAC is called following a power law, if $\psi(t) \propto t^\tau$ for $0 < \tau \leq 1$. From Table \ref{ta}, the SAD at time $t$ for a power law has the form
\[
P(X=k) = \frac{(1-\beta) (\beta)_{k-1}}{k!},~~~(k =1, 2, \ldots)
\]
where $0 \leq \beta < 1$ ($\beta$ is $1/c$ in Table \ref{ta}), and $(a)_i=a(a+1)\ldots (a+i-1)$ for $i \geq 1$ and $(a)_0=1$ is the rising factorial (this distribution appears in \citet{zhou2017frequency}). This SAD distribution does not depend on $t$. We call this discrete distribution, the {\em power species abundance distribution} (PSAD). If $X$ follows a PSAD with parameter $\beta$, then $X-1$ follows the generalized hypergeometric distribution, $_2F_1(\beta,1;2;1)$ distribution \citep{kemp1956,kemp1968}.

\vspace{0.3cm}
\noindent
{\bf Proposition 3:} {\em Under the MPPP, power law is the only ESAC which has SAD independent on $t$. Furthermore, if an SAD under MPPP has a proper limiting distribution when $t$ approaches infinity, then the limiting distribution must be a PSAD.}

\vspace{0.3cm}
The proof of Proposition 3 is given in Appendix \ref{AppD}.

Three distributions in $\mathrm{LDR}_1$ are exceptional. They stand for three boundary situations. The first one is the zero-truncated Poisson distribution ($c=0$). It models the extreme case when all species have equal abundance. In the approach that assumes finite $D$, it is a common reference distribution, say in interpreting $^q\!D_\nu$ and deriving nonparametric estimator of $D$.

The second one is the log-series distribution ($c=1$). It is where $E(D)$ jumps from finite value to infinite value in $\mathrm{LDR}_1$. Therefore, if we are interested only in finite $D$ models, it is a boundary case. Because of this property, it is used in this paper as a reference distribution in graphical check for the finiteness of $E(D)$ (refer to the end of Section \ref{s:ESAC}, and the checking of slope 1 in the D1/D2 plot introduced later).

The last one is the power law ($b=0$). It is the only possible limiting distribution of SAD in MPPP as $t$ approaches $\infty$. All SADs in $\mathrm{LDR}_1$ with $c > 1$ converge to it when $t$ increases without bound. It is also the only distribution in $\mathrm{LDR}_1$ that has $E(S(t)) = \infty$ for any $t > 0$. 

We extend the linear first derivative ratio family to linear $j$th derivative ratio family, which we denote as $\mathrm{LDR}_j$. A $\psi(t)$ belongs to $\mathrm{LDR}_j$ if $-\psi^{(j)}(t)/\psi^{(j+1)}(t)$ is a linear function of $t$. We prove in Appendix \ref{AppE} that $\mathrm{LDR}_2 = \mathrm{LDR}_3 = \ldots$, and $\mathrm{LDR}_2$ is simply a mixture of zero-rate species and $\mathrm{LDR}_1$ (i.e., the $\tilde \nu$ of $\mathrm{LDR}_2$ satisfies (\ref{eq:nu2}), but $\nu(\{0\})$ can be positive). 

\section{Diagnostic Plots} \label{s:Diag}

An advantage of $\mathrm{LDR}_1$ is that it has a simple diagnostic plot: Draw $\hat \xi(t) = -\hat{\psi}^{(1)}(t)/\hat{\psi}^{(2)}(t)$
as a function of $t\in [0,t_{0}]$ for $\hat{\psi}^{(1)}(t)$ and $\hat{\psi}^{(2)}(t)$
defined in \eqref{eq:dpsi}. If
the curve in the plot is almost linear, $\mathrm{LDR}_1$ is an appropriate model. 
 Approximate intercept and slope of the curve  can be used as initial estimate of $b$ and $c$ in finding the MLE. We call the plot, D1/D2 plot, and the curve for $\hat \xi(t)$ in the plot, the D1/D2 curve.

Similarly, to investigate how well $\mathrm{LDR}_{2}$ fits a data, we can plot the function $-\hat{\psi}^{(2)}(t)/\hat{\psi}^{(3)}(t)$ for $t\in [0,t_{0}]$
where $\hat{\psi}^{(2)}(t)$ and $\hat{\psi}^{(3)}(t)$ are defined
in \eqref{eq:dpsi}. We call the plot, D2/D3 plot, and the curve in it, the D2/D3 curve. 

As $N_1(t_0), N_2(t_0), \ldots$ are independent and Poisson distributed, by the delta method, we can approximate $Var(\hat \xi(t))$ by
\[
\widehat{Var} ( \hat \xi(t)) = \frac{\widehat{Var}(\hat \psi^{(1)}(t))}{\hat \psi^{(2)2}(t)} + \frac{\hat \psi^{(1)2}(t) \widehat{Var}(\hat \psi^{(2)}(t))}{\hat \psi^{(2)4}(t)} - \frac{2\hat \psi^{(1)}(t) \widehat{Cov}(\hat \psi^{(1)}(t),\hat \psi^{(2)}(t))}{\hat \psi^{(2)3}(t)},
\]
where
\begin{equation}
\widehat{Var}(\hat \psi^{(1)}(t)) = \frac{1}{t_0^2}\sum_{k=1}^\infty k^2 N_k(t_0)(1-t/t_0)^{2k-2}, \label{eq:vard1}
\end{equation}
\[
\widehat{Var}(\hat \psi^{(2)}(t)) = \frac{1}{t_0^4}\sum_{k=2}^\infty k^2 (k-1)^2 N_k(t_0)(1-t/t_0)^{2k-4},
\]
and 
\[
\widehat{Cov}(\hat \psi^{(1)}(t), \hat \psi^{(2)}(t)) = -\frac{1}{t_0^3}\sum_{k=2}^\infty k^2 (k-1)N_k(t_0)(1-t/t_0)^{2k-3}.
\]
We use $\hat \xi(t) \pm 1.96 \sqrt{\widehat{Var}(\hat \xi(t))}$ as an approximate 95\% pointwise confidence band for $\hat \xi(t)$ in the D1/D2 plot. Similar confidence band can be constructed in D2/D3 plot (see Appendix \ref{AppF}).

The diagnostic checks in Table \ref{ta} suggest graphs to examine the fitness of zero-truncated Poisson distribution, geometric distribution, log-series distribution, and power law to the FoF. Graphical check for the leading three distributions exist in the statistical literature. See for example \citet{hoaglin1985}. Our plots are new additions from a totally new point of view. The plot focuses on $\hat \psi^{(1)}(t)$, the estimated expected total rate of unseen species at time $t$. The graphical check detects discrepancy between the estimated function and its expected pattern for $t \in [0,t_0]$ when a specified distribution is assumed. Since the plotted $y$-value in the diagnostic plot has the form $g(\hat \psi^{(1)}(t))$, an approximate pointwise confidence band for the curve can be obtained using the approximation $Var(g(\hat \psi^{(1)}(t))) \approx [g^{(1)}(\hat \psi^{(1)}(t))]^2 \widehat{Var}(\hat \psi^{(1)}(t))$ for $\widehat{Var}(\hat \psi^{(1)}(t))$ defined in (\ref{eq:vard1}).

Currently a standard diagnostic plot for power law is the log-log plot which plots $\log(\hat \psi(t))$ against $\log(t)$ for $0 \leq t \leq t_0$.  As $d \log(\psi(t)) / d \log(t) = p_1(t)$,
log-log plot detects whether $p_1(t)$ is a constant function. As suggested by the diagnostic check of power law in Table \ref{ta}, we can plot $\log(\hat \psi^{(1)}(t))$ against $\log(t)$ for $0 \leq t \leq t_0$. We call it log(D)-log plot. Log(D)-log plot checks whether $p_2(t)/p_1(t)$ is a constant function because $d \log(\psi^{(1)}(t)) / d \log(t) = -2p_2(t)/p_1(t)$.
Log(D)-log plot is more sensitive to discrepancies with the power law because $p_1(t)$ changes very slowly with respect to $t$ for many SADs. 
It is well-known in species-area relationship studies that the curve in log-log plot is approximately linear for various dissimilar SADs \citep{preston1960time,preston1962,may1975patterns,martin2006}. 

\begin{figure}
\begin{center}
\includegraphics[width=12cm,height=16.5cm]{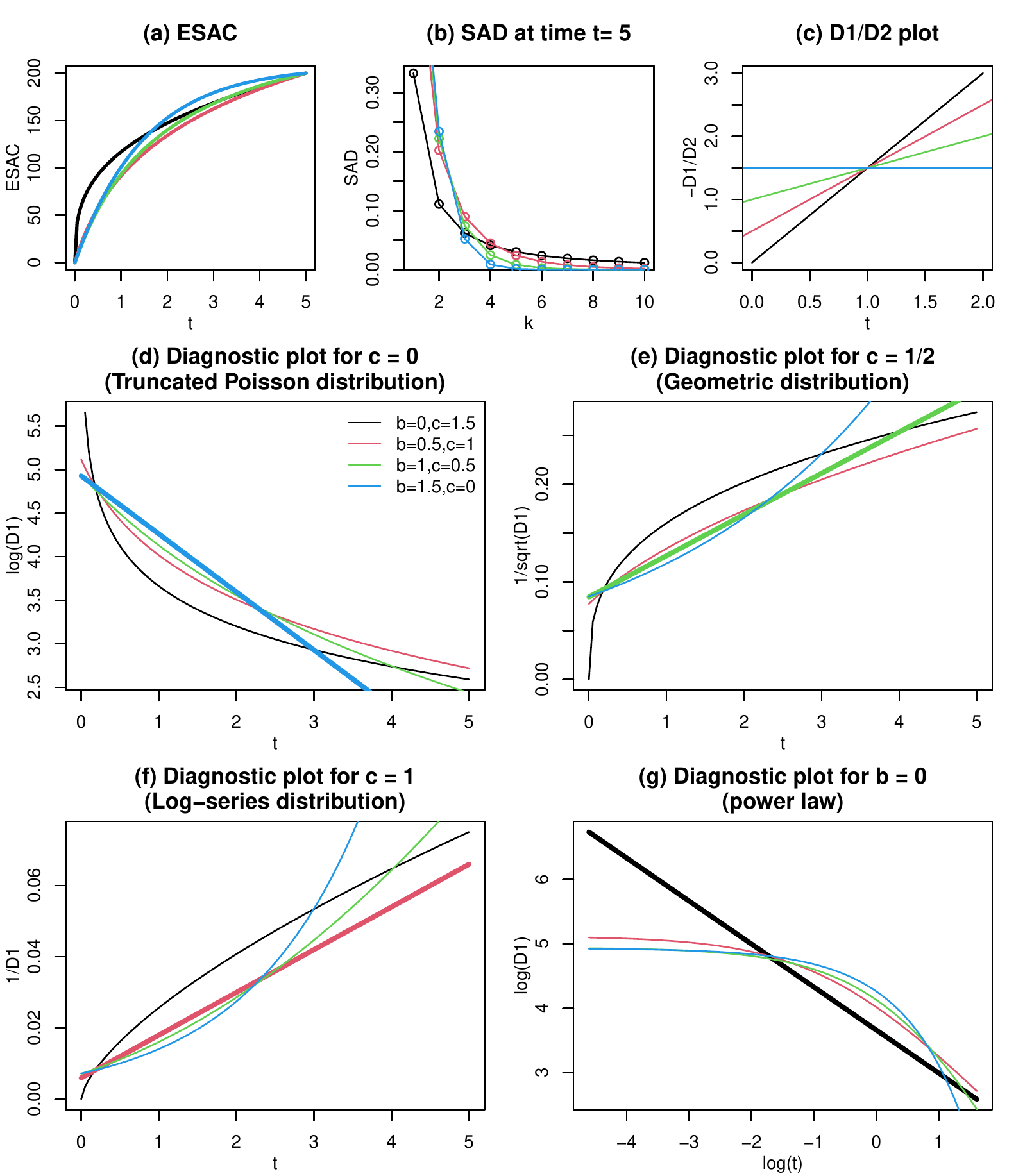}
\end{center}
\caption{Plots for four $\mathrm{LDR}_1$ models with $\xi(t) = (1.5-c)+ct$ for $c = 1.5, 1, 0.5, 1$. When $c=1.5$ ($b=0$), it is a power law and the curve is shown in black color. When $c=1$, it is a log-series distribution, and the curve is shown in red. When $c=0.5$, it is a geometric distribution, and the curve is shown in green. When $c=0$, it is a zero truncated Poisson distribution, and the curve is shown in blue. In panel (a), we draw the ESAC. In panel (b), we draw SAD at time $t=5$. In panel (c), we draw D1/D2 plot. In panels (d)-(g), we draw the graphical checks for zero truncated Poisson distribution, geometric distribution, log-series distribution and power law respectively. The straight line in the plots are drawn with larger weight. }
\label{f:LDR1}
\end{figure}

In Figure \ref{f:LDR1}, we consider four $\mathrm{LDR}_1$ distributions for which the diagnostic graphs are designed. The parameter $(b,c)$ are (1.5, 0) (zero-truncated Poisson distribution), (1, 0.5) (geometric distribution), (0.5, 1) (log-series distribution) and (0, 1.5) (power law). Parameter $a$ is chosen so that $\psi(5)=200$. We draw the ESAC in panel (a) and the SAD at $t=5$ in panel (b). We can discriminate the power law (the black curve in the panel (a)) from other distributions as its $\psi^{(1)}(0) = \infty$ ($\psi^{(1)}(0) = \Lambda$ is the expected total rate). Other than this, we learn little about the parameters $b$ and $c$ from visual inspection of the plots in panels (a) and (b). We need special plots to discriminate different parameter values. In D1/D2 plot in panel (c), the four distributions correspond to four straight lines with different slope $c$. The graphical checks in panels (d)-(g) are designed for each special distribution so that when FoF follows that distribution, the curve in the plot is a straight line, which is drawn as a heavy line in Figure \ref{f:LDR1}. Therefore, how straight the curve is can be used to assess how well the distribution fits the data. 

\section{Species Richness via Extrapolation} \label{s:Local}

Assuming the whole curve $\xi(t)$ to be linear may be too ambitious. If species richness is our main concern, our aim is to estimate $E(N_0(t_0))$. It is convenient to concentrate only on the rare species. We classify species into two groups: rare species and abundant species. Assume that (B1) all species represented 0, 1, 2, and 3 times in time interval $[0,t_0]$ are rare, and (B2) rare species follows   $\mathrm{LDR}_1$ distribution. \footnote{The categorization of species into rare species and abundant species in (B1) and (B2) is vague and data-dependent. A clear-cut threshold at 3 for the abundant species is artificial. It leads to an intuitive but debatable conclusion: All abundant species are represented at least four times in $[0,t_0]$. Results deduced from (B1) and (B2) can only be approximations. } 
Observations $n_1(t_0), n_2(t_0)$ and $n_3(t_0)$ are data from this model truncated at both ends. We do not know how many rare species are represented four times or more in $[0,t_0]$ because the observed $n_j(t_0)$ for $j \geq 4$ may include abundant species. 

It can be shown that for $\mathrm{LDR}_1$,  
\begin{equation}
\frac{E(N_j(t_0))E(N_{j+2}(t_0))}{E^2(N_{j+1}(t_0))} = \frac{(j+1)(jc +1)}{(j+2)((j-1)c+1)}, \label{eq:ratio}
\end{equation}
when $j=1, 2, \ldots$ and is independent on $t_0$. When $j=1$, we have
\[
\frac{E(N_1(t_0)) E(N_3(t_0))}{E^2(N_2(t_0))} = \frac{2(c+1)}{3}. 
\]
The information about the rare species from $N_1(t_0), N_2(t_0)$ and $N_3(t_0)$ is barely enough to make $c$ estimable. When $N_2(t_0) > 0$, a plug-in estimator of $c$ is 
\[
\hat c^* = \frac{3 N_1(t_0) N_3(t_0)}{2 N_2^2(t_0)}-1. 
\]
Estimator $\hat c^*$ needs modification to take into account two inequality constraints:  $c \geq 0$ and $b \geq 0$. Under $\mathrm{LDR}_1$, the constraint $b \geq 0$ is equivalent to $c$, the slope of $\xi(t)$ satisfying $E(N_1(t_0))/(2E(N_2(t_0))) = \xi(t_0)/t_0 \geq c$.  Therefore, our final estimator of $c$ when $N_2(t_0)>0$ is
\[
\hat c_F = \max(\min(\hat c^*,N_1(t_0)/(2N_2(t_0))), 0).
\]
If $0 \leq c < 1$, $E(N_0(t_0))$ ($=E(D)-\psi(t_0)$) is finite, and (\ref{eq:ratio}) remains true when $j=0$. We have  
\begin{equation}
E(N_0(t_0)) = \left\{ \begin{array}{ll} E^2(N_{1}(t_0))/[2(1-c)E(N_2(t_0))] & (0 \leq c < 1) \\ \infty & (c \geq 1). \end{array} \right. \label{eq:ED1}
\end{equation}
Using estimator $\hat c_F$ and (\ref{eq:ED1}), we have the following estimator of $E(N_0(t_0))$. 
\begin{equation}
\hat E(N_0(t_0)) = \left\{ \begin{array}{ll}  N_1^2(t_0)/[2(1-\hat c_F)N_2(t_0)] & (N_2(t_0)>0, 0 \leq \hat c_F <1) \\ 
\infty & (N_2(t_0) > 0, \hat c_F \geq 1) \mbox{ or } \\ & ~~~~~(N_1(t_0)>0, N_2(t_0) = 0) \\ 
0 & (N_1(t_0) = N_2(t_0) = 0) \label{eq:unseen}
. \end{array} \right. 
\end{equation} 
From (B1), all unseen species are rare. Estimator $\hat E(N_0(t_0))$ depends only on rare species data $N_1(t_0), N_2(t_0)$, and $N_3(t_0)$. When $\hat c_F = 0$, $\hat E(N_0(t_0))$ reduces to Chao1 estimator \citep{chao1984} of $E(N_0(t_0))$, which is  $N_1^2(t_0)/[2N_2(t_0)]$. If $N_2(t_0)>0$ and $\hat c_F \geq 1$, a reasonable estimate of $E(N_0(t_0))$ is infinity. When $N_2(t_0) = 0$, our estimate of $E(N_0(t_0))$ is $\infty$ when $N_1(t_0)>0$, and is 0 when $N_1(t_0)=0$. From (B1), all abundant species are observed before time $t_0$. Their contribution to $D$ is included in $N_+(t_0)$. 
Our estimator of $E(D)$ is $\hat E^*(D) = N_+(t_0)+ \hat E(N_0(t_0))$.

Chao1 estimator \citep{chao1984} is a lower bound estimator of $D$, and works well when the rare species have equal abundance, which corresponds to the $\mathrm{LDR}_1$ model with $c=0$. Our estimator $\hat E^*(D)$ is an extension of Chao1 estimator in the sense that we assume the rare species follow $\mathrm{LDR}_1$ model, and estimate the parameter $c$ using the FoF of rare species. It reduces to Chao1 estimator when $\hat c_F = 0$. 

A better way to understand the estimator $\hat E^*(D)$ is to relate it to an extrapolation of $\hat \xi(t)$ because we can assess its suitability in the D1/D2 plot. Abundant species usually appear early. Species that show up late are likely rare. The rear part of $\hat \xi(t)$ depends mainly on rare species. Extrapolation is a technique to extend our knowledge about rare species to the unseen species. (B1) and (B2) imply that $\xi(t)$ is approximately linear when $t \geq t_0$.\footnote{We do not require $\xi(t)$ to be exactly linear when $t \geq t_0$ because given the rear part of $\xi(t)$, we can find $E(N_i(t))$ for $i \geq 1$ and $t \geq t_0$ and then perform interpolation using Equation (\ref{eq:iter}) to find $\xi(t)$ for all $t \geq 0$. If $\xi(t)$ is linear after $t_0$, it can be shown that $\xi(t)$ is linear in $[0,\infty)$.} Consider two simple linear extrapolation methods in D1/D2 plot. The zeroth-order extrapolation uses $\hat \xi(t) = \hat \xi(t_0) = t_0N_1(t_0)/[2N_2(t_0)]$ for $t > t_0$.  Another extrapolation uses 
\begin{equation}
\hat \xi(t) = \hat \xi(t_0) + \hat c_F (t-t_0),~~~(t > t_0). \label{eq:cFext}
\end{equation}
As $\hat c^* =\hat \xi^{(1)}(t_0)$, (\ref{eq:cFext}) is the first-order extrapolation of $\hat \xi(t)$ when $\hat c_F = \hat c^*$. We call (\ref{eq:cFext}) a modified first-order extrapolation. It is always true that
\begin{equation}
\psi(t) = \psi(t_0)+\psi^{(1)}(t_0)\int_{t_0}^t  \exp \left( - \int_{t_0}^y \frac{1}{\xi(x)}dx \right) dy,~~(t \geq t_0). \label{eq:extra}
\end{equation}
If we estimate $\psi(t_0)$ by its MLE $N_+(t_0)$ and $\psi^{(1)}(t_0) = E(N_1(t_0))/t_0$ by its plug-in estimator $N_1(t_0)/t_0$, then we can estimate $\psi(t)$ for all $t > t_0$ using (\ref{eq:extra}) once an extrapolation rule for $\xi(t)$ is chosen. Species richness is just the limiting value of this $\psi(t)$. If zeroth-order extrapolation is used, the estimator of $E(D)$ is Chao1 estimator. If modified first-order extrapolation rule is used, the estimator of $E(D)$ is $\hat E^*(D)$. 
As $\xi(t)$ is nondecreasing, zeroth-order extrapolation is a lower bound of all reasonable extrapolations. It explains why Chao1 estimator estimates a lower bound of $E(D)$. 

All $\xi(t)$'s that fulfill the sufficient condition in Proposition 2 are asymptotically linear because its derivative is nonnegative and non-increasing and thus must have a finite limit. Furthermore, for concave $\xi(t)$, the modified first-order extrapolation is probably the first-order extrapolation. Therefore, under the sufficient condition in Proposition 2, modified first-order extrapolation is an adequate extrapolation curve when $t_0$ is large, and $\hat E^*(D)$ should give a plausible estimate.

Equation (\ref{eq:extra}) is useful. If $\xi(t) \geq \beta+t$ for a $\beta > 0$ when $t \geq t^* \geq 0$, then 
\begin{eqnarray*}
\psi(t) & \geq & \psi(t^*) + \psi^{(1)}(t^*) \int_{t^*}^t  \exp \left( - \int_{t^*}^y \frac{1}{\beta +x}dx \right) dy \\
& = & \psi(t^*) + \psi^{(1)}(t^*) (\beta +t^*) \log \left( \frac{\beta + t}{\beta + t^*}\right) \to \infty ~~\mbox{when } t\to \infty.
\end{eqnarray*} 
Similarly, if $\xi(t) < \beta+ \gamma t$ for a $\beta > 0$ and $0 \leq \gamma <1$ when $t \geq t^* \geq 0$, then 
\begin{eqnarray*}
\psi(t) & < & \psi(t^*) + \psi^{(1)}(t^*) \int_{t^*}^t  \exp \left( - \int_{t^*}^y \frac{1}{\beta + \gamma x}dx \right) dy \\
& = & \psi(t^*) + \psi^{(1)}(t^*) \frac{(\beta +\gamma t^*)^{1/\gamma}}{1-\gamma} \left( (\beta + \gamma t^*)^{1-1/\gamma}-(\beta + \gamma t)^{1-1/\gamma} \right) 
\end{eqnarray*}
where the last expression has finite limit when $t$ increases to infinity.
These two facts can be used to detect the finiteness of $E(D)$ in D1/D2 plot. If we can judge from the $\hat \xi(t)$ in the D1/D2 plot that there are positive values $\beta$ and $t^*$ such that $\hat \xi(t) \geq \beta+t$ whenever $t \geq t^*$, then $E(D)$ is likely infinite. On the other hand, if $\hat \xi(t) \leq \beta + \gamma t$ for a $0 \leq \gamma < 1$, it is reasonable to believe that $E(D)$ is finite. To assist judging whether the rear part of $\hat \xi(t)$ has slope larger than or smaller than 1, it is helpful to add grid lines  with slope 1 in the D1/D2 plot as demonstrated in Figure \ref{f:D1D2}.

We can assess the adequacy of Chao1 estimator through investigating how well $N_1(t_0)$, $N_2(t_0)$ and $N_3(t_0)$ fit a truncated Poisson distribution by testing 
\noindent
\[
H_0: 3E(N_1(t_0))E(N_3(t_0)) - 2 E^2(N_2(t_0)) \leq 0
\]
against
\[
H_1: 3E(N_1(t_0))E(N_3(t_0)) - 2 E^2(N_2(t_0)) > 0.
\]
If the null hypothesis is not rejected, Chao1 estimator gives reasonable estimate of $E(D)$, otherwise it only estimates a lower bound of $E(D)$. An unbiased estimator of $3E(N_1(t_0))E(N_3(t_0))-2E^2(N_2(t_0))$ is $T=3N_1(t_0)N_3(t_0)-2N_2(t_0)(N_2(t_0)-1)$ which is our test statistic. An unbiased estimator of $Var(T)$ is
\[
\widehat{Var}(T) = 9N_1(t_0)N_3(t_0)(N_1(t_0)+N_3(t_0)-1) + 8N_2(t_0)(3-5N_2(t_0)+2N_2^2(t_0)).
\]
The approximate p-value of the test is $P(N(0,1) \geq T/\sqrt{\widehat{Var}(T)}~)$, where $N(0,1)$ stands for standard normal distribution. We reject $H_0$ at  significance level $\alpha$ when the p-value is less than $\alpha$.

\section{Rational First Derivative Ratio Family} \label{s:extend}

\noindent
When the curve in the D1/D2 plot is not linear, but concave, it is natural to model $\xi(t)$ as a rational function. Quotient of two linear functions is too restrictive because it is in general asymptotically flat. It not only implies that $E(D)$ is usually finite, but also that the rare species are homogeneously abundant. Therefore, we consider a ratio of a quadratic polynomial and a linear function. This form of $\xi(t)$ is asymptotically linear with flexible slope. After simple manipulation, we obtain the following expression for $\xi(t)$.

\vspace{0.3cm}
\noindent
{\bf Definition: (Rational First Derivative Ratio Family)} A first derivative ratio, $\xi(t)$ belongs to the {\em rational first derivative ratio family} (denoted as $\mathrm{RDR}_1$) if 
\begin{equation}
\xi(t) = \frac{1}{c_1/(t+b_1)+c_2/(t+b_2)}, \label{eq:rdr1}
\end{equation}
where $b_2$ and $c_1$ are positive parameters, and $b_1$ and $c_2$ are nonnegative parameters. For uniqueness we assume $b_1 < b_2$. If $b_1 = 0$, we require $c_1 < 1$. This additional restriction is to ensure that $\psi(t)$ is finite for all finite $t$. 

\vspace{0.3cm}
Function $\xi(t)$ in (\ref{eq:rdr1}) fulfills the sufficient condition in Proposition 2. When $c_2 = 0$, $\mathrm{RDR}_1$ model becomes $\mathrm{LDR}_1$ model. $\mathrm{RDR}_1$ has five parameters, $a=\psi^{(1)}(t_0)$, $b_1$, $b_2$, $c_1$, and $c_2$.

To calculate the log-likelihood function using (\ref{eq:psiL}), we need to compute $\psi(t_0)$ and $\psi^{(k)}(t_0)$ for all $k$ such that $n_k(t_0) > 0$. The following equality which can be proved by mathematical induction for $k \geq 1$ is helpful,
\[
\begin{aligned} 
& k(k-1+c_1+c_2)\psi^{(k)}(t) +[c_1t+b_2c_1+c_2t+b_1c_2+k(2t+b_1+b_2)]\psi^{(k+1)}(t) \\
& +(t+b_1)(t+b_2)\psi^{(k+2)}(t) = 0. \end{aligned}
\]
To use it, choose a value for $t_0$. Given $a = \psi^{(1)}(t_0)$ which is a scale parameter, and the values of the parameters $b_1$, $b_2$, $c_1$ and $c_2$, we can find $\psi^{(2)}(t_0) = -\psi^{(1)}(t_0)/\xi(t_0)$. Then apply the above recurrence relation for $t=t_0$ to compute all necessary $\psi^{(k)}(t_0)$ in (\ref{eq:psiL}). The ESAC is   
\[
\psi(t) = a (t_0+b_1)^{c_1}(t_0+b_2)^{c_2} \int_0^t (x+b_1)^{-c_1} (x+b_2)^{-c_2} dx. 
\]
Thus $E(D)<\infty$ if and only if $c_1 + c_2 > 1$. 
It can be proved that $\nu(\{0\})=0$. We give the expression for $^q\!D_\nu$ in Appendix \ref{HillRDR}. 

\section{Examples}
\label{s:swine}

Four real FoF data are presented in Table \ref{t1}. Nonparametric analysis of the data can be found in \citet{bohning2005,lijoi2007,wang2010,chee2016,norris1998non}, and \citet{chiu2016estimating}. In this section, we fit the data using the parametric models in Sections \ref{s:LDRF} and \ref{s:extend}, demonstrate the use of various diagnostic plots, and compare different diversity estimates. Without loss of generality, we set $t_0=1$.
The significance level of the tests is fixed to 5\%. 

{ \renewcommand{\arraystretch}{1.4}
\begin{table}
\caption{Four real FoF data and the MLE for the selected model using AIC} 
\begin{center}
\begin{tabular}{cccccccccccccc}
\hline
\multicolumn{14}{l}{(i) Swine feces data} \\
\multicolumn{14}{l}{~~~~~~($\hat a = 8025.0, \hat b_1 = 0.429, \hat b_2=3.178, \hat c_1 = 0.115, \hat c_2 = 0.294$ under $\mathrm{RDR}_1$)} \\
$k$ & 1 & 2 & 3 & 4 & 5 & 6 & 7 & 8 & 9 & 10 & 11 & & \\
$n_{k}(t_{0})$ & 8025 & 605 & 129 & 41 & 16 & 8 & 4 & 2 & 1 & 1 & 1 & & \\
\hline
\multicolumn{14}{l}{(ii) Accident data} \\
\multicolumn{14}{l}{~~~~~~($\hat a = 1318.1, \hat b_1 = 0.617, \hat b_2=198.9, \hat c_1 = 0.211, \hat c_2 = 45.362$ under $\mathrm{RDR}_1$)} \\ 
$k$ & 1 & 2 & 3 & 4 & 5 & 6 & 7 & & & & & & \\
$n_k(t_0)$ & 1317 & 239 & 42 & 14 & 4 & 4 & 1 & & & & & & \\
\hline
\multicolumn{14}{l}{(iii) Tomato flowers data} \\
\multicolumn{14}{l}{~~~~~~($\hat a = 1433.7, \hat b_1 = 0.050, \hat b_2=1.451, \hat c_1 = 0.074, \hat c_2 = 0.693$ under $\mathrm{RDR}_1$)} \\
$k$ & 1 & 2 & 3 & 4 & 5 & 6 & 7 & 8 & 9 & 10 & 11 & 12 & 13 \\
$n_k(t_0)$ & 1434 & 253 & 71 & 33 & 11 & 6 & 2 & 3 & 1 & 2 & 2 & 1 & 1 \\ 
$k$ & 14 & 16 & 23 & 27 & & & & & & & & & \\
$n_k(t_0)$ & 1 & 2 & 1 & 1 & & & & & & & & & \\
\hline
\multicolumn{14}{l}{(iv) Bird abundance data ($\hat a=14.696, \hat b=0.044, \hat c=0.772$ under $\mathrm{LDR}_1$)} \\
$k$ & 1 & 2 & 3 & 4 & 5 & 6 & 7 & 8 & 9 & 10 & 12 & 13 & 14 \\
$n_k(t_0)$ & 11 & 12 & 10 & 6 & 2 & 5 & 1 & 3 & 2 & 4 &  1 & 1 & 1 \\
$k$ & 15 & 16 & 18 & 25 & 29 & 30 & 32 & 39 & 44 & 53 & 54 & & \\
$n_k(t_0)$ & 2 & 1 & 2 & 1 & 1 & 1 & 1 & 1 & 1 & 1 & 1 & & \\
\hline
\end{tabular}
\label{t1}
\end{center}
\end{table}
}

The first data is the swine feces data which appeared and was analyzed in \citet{chiu2016estimating}. It is for the pooled contig spectra from seven non-medicated
swine feces. The large $n_{1}(t_{0})$ relative to other frequencies is
viewed as a signal for sequencing errors. \citet{chiu2016estimating}
proposed a nonparametric estimate of the singleton count basing on the other counts, and the difference of this estimate and the observed singleton count is interpreted as outcome of missequencing. An implicit assumption of this approach is that sequencing errors inflate only the singleton count, and all other frequency counts are unaffected. It is equivalent to claim that there are sequencing errors which create solely zero-rate species. 

\begin{figure}
\begin{center}
\includegraphics[width=5in]{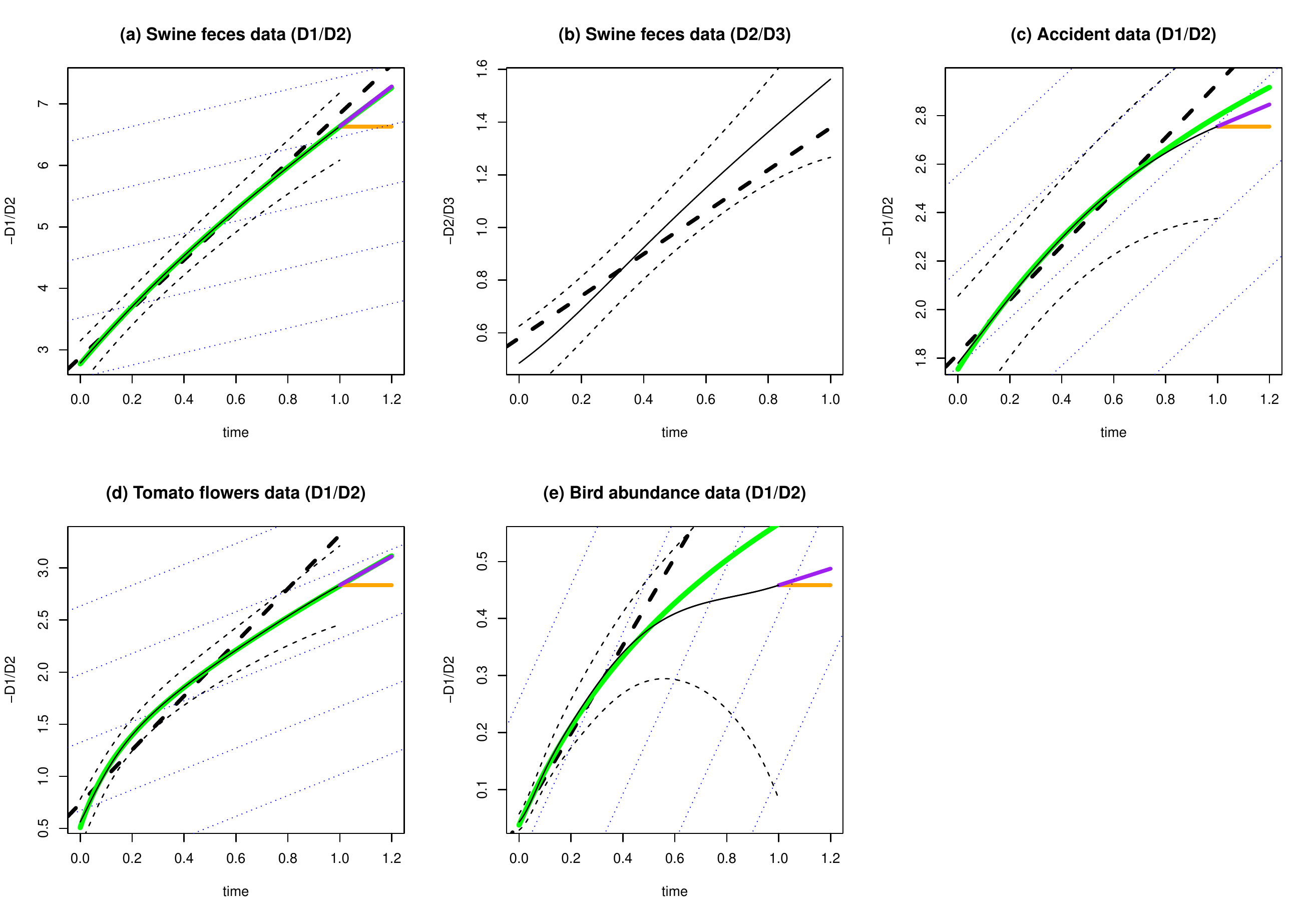}
\end{center}
\caption{D1/D2 and D2/D3 plots for the data. Panels (a), (c), (d) and (e) are D1/D2 plots for the four examples, and Panel (b) is the D2/D3 plot for the first example. The light black solid curves are the D1/D2 or D2/D3 curve. The light black dashed curves are the 95\% pointwise confidence bands. The heavy black dashed lines are the lines fitted by the MLE under $\mathrm{LDR}_1$. The heavy green solid curves in panels (a), (c), (d) and (e) are the fitted curves under $\mathrm{RDR}_1$. The blue dotted lines in the D1/D2 plots are grid lines with slope 1. They are added to assist checking the slope of the curve. D1/D2 curve with slope larger than 1 is a signal for infinite $E(D)$, while slope less than 1 is for finite $E(D)$. 
The zeroth-order and the modified first-order extrapolation lines are drawn in D1/D2 plot in orange and purple respectively. They correspond to the extrapolation used by Chao1 estimator (orange line) and $\hat E^*(D)$ (purple line) respectively.} 
\label{f:D1D2}
\end{figure}

To investigate whether zero-rate species really exist, we draw the D1/D2 and D2/D3 plots with 95\% pointwise confidence bands in panels (a) and (b) respectively in Figure \ref{f:D1D2}. The approximate linear curve in both plots indicates that both $\mathrm{LDR}_1$ and $\mathrm{LDR}_2$ are reasonable models. 
The heavy dashed lines in panels (a) and (b) are the lines fitted by MLE under $\mathrm{LDR}_1$. The fitted line is close to the curve in D1/D2 plot, but is not so in the D2/D3 plot. As the dashed line lies inside the confidence bands in panel (b), the disagreement between the singleton count and other counts is not strong enough to reject that they come from the fitted $\mathrm{LDR}_1$. The Pearson's chi-square test statistic after grouping all cells with expected frequency less than 5 is 4.325 with 4 degrees of freedom.
The estimated $E(N_1(t_0))$ under this $\mathrm{LDR}_1$ is 8027.6 which is marginally larger than $n_1(t_0)$. As $\hat E(N_1(t_0)) > n_1(t_0)$, the MLE of $\nu(\{0\})$ under $\mathrm{LDR}_2$ should be zero. There is no significant evidence for the existence of zero-rate species. 
The heavy green curve in panel (a) is the fitted curve under $\mathrm{RDR}_1$. The heavy green curve is very close to the D1/D2 curve. $\mathrm{RDR}_1$ performs better than $\mathrm{LDR}_1$ in terms of Akaike information criterion (AIC) (the AIC for $\mathrm{LDR}_1$ is -136060.6, and that for $\mathrm{RDR}_1$ is -136061). \footnote{The standard likelihood ratio test is not valid as the tested parameter value is on the boundary of the parameter space.}

In panel (a) blue grid lines with slope 1 are drawn. From Section \ref{s:Local}, the slope of the rear part of $\hat \xi(t)$ larger (or smaller) than 1 is an indication that $E(D)$ is infinite (or finite). Apparently the slope of $\hat \xi(t)$ has slope always larger than 1. We are confident that $E(D)$ is infinite. The modified first-order extrapolation line for $\hat E^*(D)$ (the purple line) and the zeroth-order extrapolation line for Chao1 estimator (the orange line) are drawn. As $\hat c_F > 1$, $\hat E^*(D)=\infty$. The orange extrapolation line does not look good.  Chao1 estimator does not perform satisfactorily.

The second data come from 9461 accident insurance policies issued by an insurance company. It was used in \citet{bohning2005,wang2010}, and \citet{chee2016}. The species corresponds to the policies, and the frequency count to the number of claims during a particular year. The $\hat \xi(t)$ in the D1/D2 plot for the accident data in panel (c) is concave but not far from linear. The heavy dashed line is the fitted line under $\mathrm{LDR}_1$. The p-value for the Pearson chi-square test for $\mathrm{LDR}_1$ is 0.0979. The MLE of $c$ is 1.1146 which is larger than 1. Therefore, the MLE of $E(D)$ is infinite. 
Compared with the blue grid lines with slope 1, the average slope of the D1/D2 curve is close to one showing uncertainty about the finiteness of  $E(D)$.  
The heavy green curve in panel (c) is the fitted curve under $\mathrm{RDR}_1$. It is preferred to $\mathrm{LDR}_1$ because it has smaller AIC (the AIC for $\mathrm{LDR}_1$ and $\mathrm{RDR}_1$ are -18691.95 and -18692.15 respectively). The MLE of $E(D)$ under $\mathrm{RDR}_1$ is 6354. It is less than the true value $D=9461$, and is comparable to the nonparametric estimates presented in \citet{chee2016} which ranges from 4016 to 7374. The purple line is the modified first-order extrapolation used by $\hat E^*(D)$. As the slope of the purple line is less than one, $\hat E^*(D)$ is finite. For this data, $\hat c_F=0.4525$ and $\hat E^*(D)=8249.2$ which is closer to the true value. The orange extrapolation line for Chao1 estimator is also drawn. The difference of the two extrapolation lines is not small. 

The third data come from a CDNA library of the expressed sequence tags of tomato flowers. It was studied in \citet{bohning2005} and \citet{lijoi2007}. The D1/D2 plot in panel (d) shows that $\mathrm{LDR}_1$ model does not fit the data well (the p-value of the Pearson chi-square test is 0.0059). The curve looks like a Bernstein function. An $\mathrm{RDR}_1$ model is fitted and the fitted curve is shown in the plot by a heavy green curve. The model fits the data well (Pearson chi-square statistic after grouping all cells with expected frequency less than 5 is 1.470 with 2 degrees of freedom). 
As $\hat c_1 + \hat c_2 = 0.767 <1$, the estimated $E(D)$ is infinite. This estimate of $E(D)$ agrees with the finding when we compare the slope of the rear part of the curve with one under the assistance of the blue grid lines. The modified first-order and zeroth-order extrapolation lines are drawn in purple and orange respectively. The difference of the slopes of the lines is not small.

The fourth data is the bird abundance data for the Wisconsin route of the North American Breeding Bird Survey for 1995. Totally 645 birds from 72 species are recorded. 
The data was studied in \citet{norris1998non} 
where a mixture of five Poisson models was fitted, and the estimated $D$ is 76. Comparing with the blue grid lines, the slope of the D1/D2 curve is less than one. Species richness should be finite. The $\hat \xi(t)$ in the D1/D2 plot does not look like a straight line. However, the $\mathrm{LDR}_1$ model is not rejected as the p-value of the Pearson's chi-square test is 0.116. The estimated $E(D)$ is 124.50. 
We also fit a $\mathrm{RDR}_1$ model to the data. The fitted line for this model is shown in the figure as a heavy green curve. The MLE of $E(D)$ under $\mathrm{RDR}_1$ is 80.7. $\mathrm{LDR}_1$ is preferred to $\mathrm{RDR}_1$ as its AIC is -25.63 which is less than that for $\mathrm{RDR}_1$ which is -24.53. $\mathrm{RDR}_1$ does not fit the rear part of $\hat \xi(t)$ well because of an abrupt change of slope around $t=0.5$. After $t=0.6$, $\hat \xi(t)$ looks quite linear. It suggests that $\hat E^*(D)$ may be a good estimate. For this data, $\hat E^*(D)=77.90$, which is close to the Chao1 estimate 77.04. 

\begin{figure}
\begin{center}
\includegraphics[width=12cm,height=14cm]{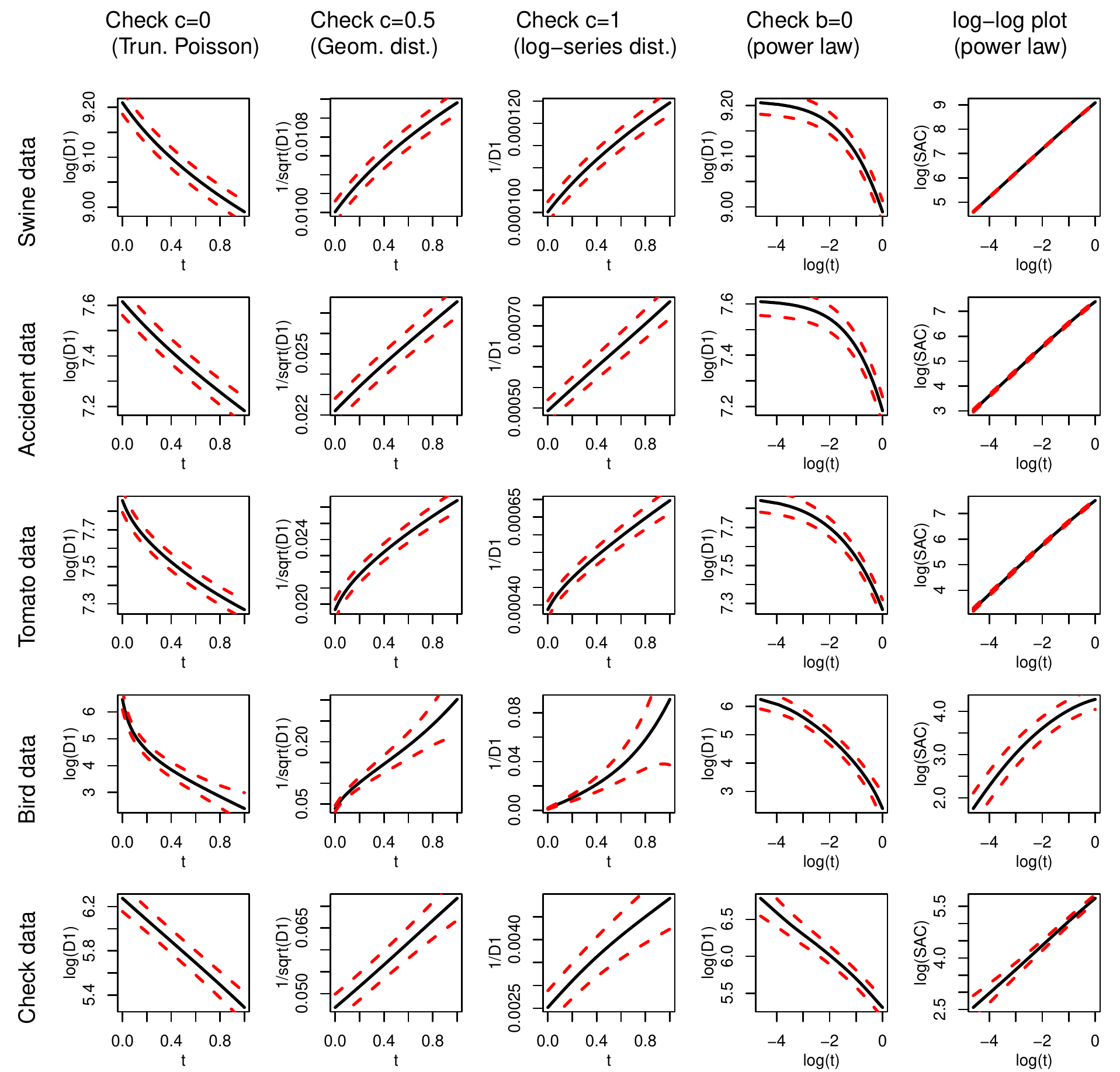}
\end{center}
\caption{Diagnostic plots for data. Each column corresponds to one diagnostic plot. The first four columns are plots suggested in Table \ref{ta}. They check $c=0$ (zero-truncated Poisson distribution), $c=0.5$ (geometric distribution), $c=1$ (log-series distribution) and $b=0$ (log(D)-log plot for power law). The last column is the log-log plot used to check the power law. The two red dashed curves are the 95\% approximate pointwise confidence band. There are five rows. The first four rows are for the real data: swine feces data, accident data, tomato flowers data and bird abundance data respectively. The last row is for data simulated from the $\mathrm{LDR}_1$ distribution to be checked. Their parameter vector $(a,b,c)$ are (200,1,0) (zero-truncated Poisson distribution), (200,1,0.5) (geometric distribution), (200,1,1) (log-series distribution), (200,0,3) (power law), and (200,0,3) (power law) respectively. When the curve is close to a straight line, it means that the checked distribution fits the data.}
\label{f:diag4}
\end{figure}

We investigate the performance of various special graphical checks in Figure \ref{f:diag4}. If the curve in the plot is close to a straight line, we assume that the corresponding distribution fits the data. The red dashed curves are the 95\% approximate pointwise confidence band. The five columns correspond to five graphical checks: namely the plot for zero-truncated Poisson distribution (check $c=0$), for geometric distribution (check $c=0.5$), for log-series distribution (check $c=1$), for power law (check $b=0$; it is the log(D)-log plot), and the log-log plot for power law. The first four rows of plots are for the four real FoF data. The last row is for data simulated from the specified distribution which the diagnostic plot is designed to check. The curves in the last row are close to a straight line as expected. The log-log plot in the last column does not perform satisfactorily. The first three graphs shows almost perfect line.  
Comparatively, the log(D)-log plot correctly declare discrepancy of power law to the four real data. The graphical check for log-series distribution correctly detects that accident data follow closely to a log-series distribution (the MLE of $c$ under $\mathrm{LDR}_1$ is 1.1146). The curves for swine feces data and the tomato flowers data are concave. It implies that $E(D)$ is infinite as pointed out at the end of Section \ref{s:ESAC}. The plot for geometric distribution and zero-truncated Poisson plot look reasonable except for the accident data. 

Consider estimation of Hill numbers. In Table \ref{t3}, we compare the model-based estimator developed in the paper, and Chao-Jost estimator proposed in \citet{chaojost2015}. For our model-based estimator, $95\%$ confidence intervals are constructed using parametric bootstrap method. Details are given in Appendix \ref{Appboot}. The difference between our estimators and Chao-Jost estimator is mild when $q=2$. In other situations, the difference can be large. Sometimes the two confidence intervals do not overlap. 
 The width of the 95\% confidence interval for our parametric estimator is larger than that of the corresponding interval for Chao-Jost estimator. The interval estimate for Chao-Jost estimator is obtained basing on the assumption that $D$ is finite and all unseen species have equal abundance. For the accident data, the true $D$ is contained in the model-based 95\% confidence interval, but not in that for the Chao-Jost estimator.

{ \renewcommand{\arraystretch}{1.4}
\begin{table}
\begin{threeparttable}[b]
\caption{Chao-Jost Estimates and Our Parametric Estimates of Hill Numbers for Selected Models (Values enclosed between square parentheses are 95\% confidence interval)}
\begin{center}
\begin{tabular}{llccc}
\hline
Data & Model/Method & $^0\!D_\nu$ & $^1\!D_\nu$ & $^2\!D_\nu$ \\
\hline
Swine & $\mathrm{RDR}_1$ & $\infty$ & 194649 & 27698 \\
feces & & $[105408, \infty]$ & $[62348, 407027]$ & $[23940, 31195]$ \\ 
data & Chao-Jost \tnote{1} & 62051 & 53835 & 27801 \\
& & $[59197, 64905]$ & $[51634, 56035]$ & $[26076, 29526]$ \\
\hline
Accident & $\mathrm{RDR}_1$ & 6354 & 5203 & 3557 \\
data & & $[4741, 16013]$ & $[4458, 6684]$ & $[2996, 4210]$ \\
(True $D$ & Chao-Jost & 5248 & 4788 & 3606 \\
$=9461$) & & $[4713, 5783]$ & $[4449, 5126]$ & $[3322, 3890]$ \\
\hline
Tomato & $\mathrm{RDR}_1$ & $\infty$ & 5941 & 1311 \\
flowers & & $[7849, \infty]$ & $[4054, 7741]$ & $[699, 2120]$ \\
data & Chao-Jost & 5887 & 4079 & 1450 \\
& & $[5405, 6370]$ & $[3809, 4349]$ & $[1254, 1646]$ \\
\hline
Bird & $\mathrm{LDR}_1$ & 124.51 & 43.84 & 28.43 \\
abundance & & $[63.63, 417.98]$ & $[32.82, 57.58]$ & $[19.55, 39.91]$ \\
data & Chao-Jost & 77.03 & 41.94 & 27.57 \\
& & $[61.68, 92.39]$ & $[39.35, 44.54]$ & $[24.72, 30.41]$  \\
\hline
\end{tabular}
\begin{tablenotes}

\item[1] Hill numbers estimates are the estimator derived in \citet{chaojost2015}. Point and interval estimates are computed using R package SpadeR \citep{chaohsieh2016}.

\end{tablenotes}
\label{t3}
\end{center}
\end{threeparttable}
\end{table}

Consider estimation of species richness. In Table \ref{Dhat}, we compare $\hat E^*(D)$, Chao1 estimator, iChao1 estimator  in \citet{chiu2014}, and the abundance-based coverage estimator (ACE) in \citet{chao1992} and \citet{chao1993}. We construct the $95\%$ confidence intervals for $E(D)$ basing on $\hat E^*(D)$ using parametric bootstrap method. Details are given in Appendix \ref{Appboot}. Estimator $\hat E^*(D)$ usually gives larger estimate. For the bird abundance data, all estimators give similar point estimate. For the accident data, $\hat E^*(D)$ gives the best estimate of the true $D$. For the remaining two data, the differences are huge. Our estimate is infinite, while other estimates are finite as assumed. 

The difference between $\hat E^*(D)$ and the other three methods can well be explained by the test on whether the observed $N_1(t_0)$, $N_2(t_0)$ and $N_3(t_0)$ fit a truncated Poisson distribution. The p-value of the test at the end of Section \ref{s:Local} for the bird abundance data is 0.374. For this data, all estimates are close to each others. The p-value for accident data is 0.040. The difference is significant, but not huge. The p-value of the other two data are smaller than $6 \times 10^{-6}$ implying that the deviation is huge.  The 95\% confidence interval for swine feces data is $[\infty, \infty]$, a much stronger signal when compared to the confidence interval in Table \ref{t3}. For $\hat E^*(D)$, only the 95\% confidence interval for bird abundance data has finite width. It has the lower endpoint close to the lower endpoint of other confidence intervals, but it has much larger upper endpoint. 

Estimator $\hat E^*(D)$ gives wider confidence interval than that for Chao1 estimator because it uses a modified first-order extrapolation that requires the estimation of slope, whereas zeroth-order extrapolation does not have this requirement. Nevertheless, first-order extrapolation looks more rational when compared to zeroth-order extrapolation which can only estimate a lower bound.
 
{ \renewcommand{\arraystretch}{1.4}
\begin{table}
\begin{threeparttable}[b]
\caption{Point and 95\% Interval Estimates of Species Richness}
\begin{center}
\begin{tabular}{lcccc}
\hline
Data & Estimator &  Estimate & 95\%Lower & 95\%Upper \\
\hline
Swine & $\hat E^*(D)$ & $\infty$ & $\infty$ & $\infty$ \\
feces & Chao1 \tnote{1} & 62051.3 &  57409.9 & 67136.2 \\
data & iChao1 \tnote{2} & 67615.0 &  62889.6 & 72753.5 \\
 & ACE \tnote{3} & 68827.7 & 62928.4 & 75370.3 \\
\hline
Accident & $\hat E^*(D)$ & 8249.2 & 5087.4 & $\infty$ \\
data & Chao1 & 5247.8 & 4682.7 & 5917.3 \\
(True D & iChao1 & 5966.7 & 5396.9 & 6622.6 \\
= 9461) & ACE & 5683.8 & 5031.1 & 6461.5 \\
\hline
Tomato & $\hat E^*(D)$ & $\infty$ & 18108.3 & $\infty$ \\
flowers & Chao1 & 5887.4 & 5274.4 & 6609.2 \\
data & iChao1 & 6512.3 & 5951.4 & 7149.5 \\
 & ACE & 6645.5 & 7779.2 & 11859.5 \\
\hline
Bird & $\hat E^*(D)$ & 77.9 & 72.5 & 297.0 \\
abundance & Chao1 & 77.0 & 73.3 & 92.1 \\
data & iChao1 & 77.5 & 74.6 & 83.3 \\
 & ACE & 78.7 & 74.3 & 91.7 \\
\hline
\end{tabular}
\begin{tablenotes}
\item[1] Chao1 estimate and the confidence interval are computed using R package SpadeR \citep{chaohsieh2016}.

\item[2] iChao1 estimator is an improved Chao1 estimator proposed in \citet{chiu2014}. The point and the interval estimates are computed using R package SpadeR \citep{chaohsieh2016}.

\item[3] Abundance-based coverage estimate (ACE) \citep{chao1992} and the confidence interval are computed using R package SpadeR \citep{chaohsieh2016}.

\end{tablenotes}
\label{Dhat}
\end{center}
\end{threeparttable}
\end{table}
}

\section{$\rho$-appearance Design}
\label{s:SD}

Usually we collect all available information within the survey period. Under our framework, all useful information is in FoF. If labor saving is our concern, we may neglect some minor information, say halt recording a species as soon as
its observed frequency reaches a fixed positive integer $\rho$ in the study period $[0, t_0]$. In this case,  
the observation period for each species varies. The period is short
for abundant species, and long for rare species. We call this design, the $\rho$-appearance
design. This design places more emphasis on rare species than abundant
species which is in line with the common understanding that information
about the rare species is critical when our interest is in $D$. In
bird survey, species can be identified by distant sightings or short
bursts of song. Stop recording abundant species early helps the researcher
concentrating more on the rare species. When $\rho=\infty$, we obtain
$\tilde n(t_0)$. When $\rho=1$, we record only the first appearance-time
of each seen species. It is exactly the information available in 
the empirical SAC.

We call the appearance time of the $\rho$th individual of a species in a survey, the $\rho$-appearance time of that species. Our observations are $\{n_1(t_0), \ldots, n_{\rho-1}(t_0),$ $r_1, r_2, \ldots, r_m\}$ where $r_1, \ldots, r_m$ are the observed  $\rho$-appearance times. The log-likelihood function given a realization $\{n_j(t_0)\}_{j=1,\ldots,\rho-1},$ $ \{r_i\}_{i=1,\ldots,m}$ is proved in Appendix \ref{AppH} to be
\begin{equation}
\log(\mathcal{L}(\psi | \{n_j(t_0)\}, \{r_i\})) 
= -\psi(t_{0}) + \sum_{j=1}^{\rho-1} n_j(t_0) \log(| \psi^{(j)}(t_{0}) | ) + \sum_{i=1}^m \log(| \psi^{(\rho)}(r_{i}) | ). \label{eq:likeJ}
\end{equation}
A numerical advantage of (\ref{eq:likeJ}) is that it does not require a general expression of $\psi^{(k)}(t)$ for all $k \geq 1$. 
A small simulation experiment in Appendix \ref{AppI} shows that the loss in information of $\rho$-appearance design when compared to the standard design is marked when $\rho=1$, and minor when $\rho=4$.

By the displacement theorem of Poisson process \citep{kingman1992poisson},
the $\rho$-appearance times form a Poisson process with intensity function
\begin{equation}
f_\rho(r)= \int\frac{(\lambda r)^{\rho-1}e^{-\lambda r}}{(\rho-1)!}\nu(d\lambda) =\frac{r^{\rho-1}}{(\rho-1)!}(-1)^{\rho-1}\psi^{(\rho)}(r). \label{eq:inten} 
\end{equation}
From (\ref{eq:deriv-1}), another expression for $f_\rho(r)$ is $f_\rho(r) = \rho E(N_\rho(r))/r$.
Equation (\ref{eq:inten}) gives another interpretation of $\psi^{(k)}(t)$. For example, $\psi^{(1)}(r)$ is the intensity function of the 1-appearance times (thus $\psi^{(1)}(r)/\psi(t_0)$ is the density function of the 1-appearance times of all seen species in $[0,t_0]$), and $r|\psi^{(2)}(r)|$ is the intensity function of the 2-appearance times. 

\section{Inference on Empirical Species Accumulation Curve} \label{s:curve2}
Suppose we only observe the empirical SAC, $n_+(t)$ for $t \in [0,t_0]$, or 
equivalently, the 1-appearance times of all seen species in $[0,t_0]$, say $r_1, \ldots, r_{n_+(t_0)}$ (i.e., the case 
$\rho=1$ in Section \ref{s:SD}). 
From (\ref{eq:likeJ}), the log-likelihood function is
$\log(\mathcal{L}(\psi \mid \{r_i\})) 
= \sum_{i=1}^{n_+(t_0)} \log(\psi^{(1)}(r_{i})) -\psi(t_{0})$.
If $\psi(t)$ has a free scale parameter, MLE of $\psi(t_0)$ is $n_+(t_0)$.
The MLE of $\psi(t)$ for power law has a simple closed form $n_+(t_0) (t/t_0)^z$, where $z=\min\{n_+(t_0)/\sum_i \log(t_0/r_i),1\}$. 

Given a parametric form of $\psi(t)$, a traditional approach is to fit it to the empirical SAC by linear or non-linear least-squares method. 
Two differences between the MLE approach and the curve-fitting method are noteworthy. First, the MLE of $\psi(t_{0})$ is equal to $n_{+}(t_{0})$ whenever $\psi(t)$ has a free scale parameter, and it is not the case in the curve-fitting approach. 
Second, the MLE method fits the density function to the 1-appearance times, 
while the curve-fitting method fits a function to the empirical SAC. 
The curve-fitting methods do not take the interdependence among the points in the empirical SAC into consideration. Since such interdependence is present in species-time curves and Type I species-area curves \citep{scheiner2003six}, the curve-fitting approach is theoretically flawed. Although improvements can be made in curve-fitting approach through transformation and/or adding weights, maximum likelihood approach is preferred for its proven effectiveness under the assumption that the model is correct.

In certain situations, only the values of the empirical SAC at a finite set of times are available. For example, only the cumulative number of species observed after day 1, day 2, and so on are recorded. Suppose the observed $N_+(\ell_i)$ is $n_+(\ell_i)$ for $i = 1, \ldots, m$ with $0 = \ell_0 < \ell_1 < \ldots < \ell_m = t_0$.
The log-likelihood function is
\begin{align*}
& \log(\mathcal{L}(\psi \mid \{n_+(\ell_i)\}_{i=1,\ldots,m})) \\
& = \sum_{i=1}^m (n_+(\ell_{i})- n_+(\ell_{i-1}))\log(\psi(\ell_i)-\psi(\ell_{i-1})) - n_+(t_0) \log(\psi(t_0)).
\end{align*}
In the simulation study in Appendix \ref{AppJ}, MLE has smaller root mean squared relative error in extrapolation when compared to the curve-fitting method.

The distribution function of the 1-appearance times in time interval $[0,t_0]$ is $\psi(t)/\psi(t_0)$. This fact holds in general for any nondecreasing $\psi(t)$ with $\psi(0)=0$, including 
sigmoid function such as the cumulative Weibull function. If we assume that the 1-appearance times are independent, we can perform maximum likelihood inference conditional on $n_+(t_0)$ when a parametric form of $\psi(t)$ is given. Full maximum likelihood calculation is possible when further assumption on the distribution of $N_+(t_0)$ is made. As the empirical SAC is proportional to the empirical distribution function of the 1-appearance times, statistical tools for empirical distribution function can be used. For example, we can apply the Dvoretzky-Kiefer-Wolfowitz inequality to construct confidence bands for the distribution function $\psi(t)/\psi(t_0)$.

\section{Discussion} \label{s:discuss} 

\noindent
A general framework, MPPP, is proposed for species abundance data. This framework has the following novel features: (1) it delineates the relation between two analytic tools -- SAD and SAC, and an MPPP can be characterized by either one of them; (2) it characterizes the class of possible ESACs to be the class of Bernstein functions, giving solid theoretical support to the empirical observation that SACs are nondecreasing concave functions; (3) it allows the existence of zero-rate species; (4) it contains the $\mathrm{LDR}_1$ model, a parametric model where the first derivative ratio of the ESAC is linear, which admits several graphical checks, and two generalizations, namely $\mathrm{LDR}_2$ which includes zero-rate species, and $\mathrm{RDR}_1$ model where the first derivative ratio is a rational function; (5) it admits a new species richness estimator $\hat E^*(D)$; (6) it admits a natural generalization of Hill numbers as measures of species diversity, which are well-defined even when the number of species is random and infinite; and (7) it allows inference to be performed when only the first $\rho$ appearance times of a species is recorded, which we refer as $\rho$-appearance design.

Compared to the conventional curve-fitting approach to SAC, our framework has a more solid theoretical underpinning, as an MPPP is uniquely characterized by an ESAC that is a Bernstein function. In the study of species-area curve, power law is popular. Nevertheless, under our model, power law is an extreme case. Its $E(D)$ and $E(S(t))$ for any positive $t$ are infinite, and its PSAD has a heavy right tail. Even though curve fitting can give reasonable estimates, the MPPP framework provides a parametric generative model in which maximum likelihood estimates can be obtained. \citet{matthews2014fitting} 
suggests using maximum likelihood methods instead of least-squares approaches in SAD fitting. We would like to extend their advice to ESAC fitting.

One possible future direction is to apply MPPP in a nonparametric setting, which lessens the need for model selection. Another future direction is to understand the tradeoff that arises in the $\rho$-appearance design. A small $\rho$ can reduce the sampling effort, at the expense of less accurate estimates of parameters. Investigating this tradeoff in a theoretic and empirical setting can provide guidance to the design.

\vspace{0.5cm}
\noindent
{\bf Acknowledgments}

\noindent
The authors are grateful to the associate editor and two anonymous referees for helpful suggestions that improved the presentation of this article.

\vspace{0.5cm}
\appendix

\section{Conditional Distribution of the Rate of an Individual at a Fixed Time} \label{AppFix}

Informally, if we condition on the event that there is one individual
at exact time $t$, then the rate of its species follows the
distribution $\nu/\Lambda$ if $\Lambda<\infty$.
The formal statement below involves a limit argument.

\vspace{0.3cm}
\noindent
{\bf Proposition A:}  {\em Fix $t\ge0$, and assume $\Lambda<\infty$.
Conditional on the event that there is at least one individual in
the time interval $[t,t+\epsilon]$ for $\epsilon >0$, then the probability that there
is exactly one individual in $[t,t+\epsilon]$ tends to $1$, and
the rate of its species converges in distribution to the distribution
$\nu/\Lambda$ as $\epsilon\to 0$. }

\vspace{0.3cm}
\noindent
{\em Proof.} 
Let $A$ be the event that there is at least one individual
in the time interval $[t,t+\epsilon]$, and $B$ be the event that there is exactly one individual in the time interval $[t, t+\epsilon]$ for $\epsilon > 0$. The probability that a species with rate $\lambda$ is observed during $[t,t+\epsilon]$ is $1-\exp(-\lambda\epsilon)$. By the thinning property, the rates of the positive-rate species observed during $[t,t+\epsilon]$ form a Poisson process with intensity measure
$(1-\exp(-\lambda\epsilon))\tilde{\nu}(d\lambda)$. Considering also
the zero-rate species, the rates of the species observed during $[t,t+\epsilon]$
form a Poisson process with intensity measure $(1-\exp(-\lambda\epsilon))\lambda^{-1}\nu(d\lambda)$
(where $(1-\exp(-\lambda\epsilon))\lambda^{-1}=\epsilon$ when $\lambda=0$). Event $A$ is that this Poisson process has at least one point. As $1-\exp(-x) \leq x$ when $x \geq 0$,
\begin{eqnarray*}
P(A) & = & 1- \exp \left( -\int_0^\infty (1-\exp(-\lambda \epsilon)) \lambda^{-1} \nu(d \lambda) \right) \\
& \leq & \int_0^\infty (1-\exp(-\lambda \epsilon)) \lambda^{-1} \nu(d \lambda) \leq \epsilon \Lambda.
\end{eqnarray*}
Similarly, the rates of the species observed exactly one time during $[t,t+\epsilon]$ form a Poisson process with intensity measure
\[
[\exp(-\lambda \epsilon) \lambda \epsilon] \lambda^{-1} \nu(d \lambda) = \epsilon \exp(-\lambda \epsilon) \nu(d \lambda). 
\]
Event $B$ is that this Poisson process has exactly one point. Thus
\[
P(B) = \exp \left( - \epsilon \int_0^\infty \exp(-\lambda \epsilon) \nu(d \lambda) \right) \int_0^\infty \epsilon \exp(-\lambda \epsilon) \nu(d \lambda),
\]
which is equal to $\epsilon \Lambda (1 + o(1))$ when $\epsilon \to 0$ because $\Lambda$ is finite.
Therefore,
\[
\lim_{\epsilon \to 0} P(B \mid A) = \lim_{\epsilon \to 0} \frac{P(B)}{P(A)} \geq \lim_{\epsilon \to 0} \frac{\epsilon \Lambda(1 + o(1))}{\epsilon \Lambda} = 1. 
\]
As probability cannot be larger than 1, it implies that $P(B \mid A) \to 1$ as $\epsilon \to 0$.

Given that exactly one individual appears in $[t,t+\epsilon]$, let $X$ be the rate of the species that the individual belongs and $F$ be its distribution function. For any  positive $\lambda_0$, let $C_{\lambda_0}$ be the event that $X$ is less than or equal to $\lambda_0$. Then
\begin{eqnarray*}
F(\lambda_0) & = & \frac{P(C_{\lambda_0})}{P(B)} = \frac{\exp \left( - \epsilon \int_0^{\lambda_0} \exp(-\lambda \epsilon) \nu(d \lambda) \right) \int_0^{\lambda_0} \epsilon \exp(-\lambda \epsilon)  \nu(d \lambda)}{\exp \left( - \epsilon \int_0^\infty \exp(-\lambda \epsilon) \nu(d \lambda) \right) \int_0^\infty \epsilon \exp(-\lambda \epsilon) \nu(d \lambda)}.
\end{eqnarray*}
It follows that 
\[
\lim_{\epsilon \to 0} F(\lambda_0) = \frac{\int_0^{\lambda_0} \nu(d \lambda)}{\Lambda}.
\]
Therefore, $X$ converges in distribution to the distribution $\nu/\Lambda$ as $\epsilon \to 0$.
 ~~~~~$\Box$

\vspace{0.3cm}
For the case $\Lambda=\infty$, as $\epsilon\to0$, the rate of the
species observed in $[t,t+\epsilon]$ diverges to infinity.

\vspace{0.3cm}
\noindent
{\bf Proposition B:} {\em 
Fix $t\ge 0$, and assume $\Lambda=\infty$. Conditional on the event
that there is at least one individual observed in the time interval
$[t,t+\epsilon]$ for $\epsilon > 0$, then the minimum rate among observed species converges
in probability to $\infty$ as $\epsilon\to0$. }

\vspace{0.3cm}
\noindent
{\em Proof.} 
Let $A$ be the event that there is at least one individual
in the time interval $[t,t+\epsilon]$. As in the case $\Lambda<\infty$ in Proposition A, for any fixed $\lambda_{0}\ge0$, we have
\begin{align*}
& P\left(\text{at least one individuals in \ensuremath{[t,t+\epsilon]} with rate \ensuremath{\le\lambda_{0}}}\right) \\
& = 1- \exp \left( -\int_0^{\lambda_0} (1-\exp(-\lambda \epsilon)) \lambda^{-1} \nu(d \lambda) \right) \\
& \leq \int_0^{\lambda_0} (1-\exp(-\lambda \epsilon)) \lambda^{-1} \nu(d \lambda) \\
& \le \epsilon\int_{0}^{\lambda_{0}}\nu(d\lambda) =O(\epsilon)
\end{align*}
as $\epsilon\to 0$ since $\int_{0}^{\lambda_{0}}\nu(d\lambda)< \infty$
due to $\int_{0}^{\infty} \min\{1,\lambda^{-1}\} \nu(d\lambda)<\infty$ by the definition of MPPP. On the other hand, as $1-\exp(-x) \geq x \max\{1-x/2,0\}$ when $x \geq 0$, 
\[
P(A) = P(N_+(\epsilon) > 0) = 1 - \exp(-\psi(\epsilon)) \geq \psi(\epsilon) \max \left\{ 1-\psi(\epsilon)/2, 0 \right\}.
\]
Since $\lim_{\epsilon \to 0} \psi(\epsilon)/\epsilon = \psi^{(1)}(0) = \Lambda = \infty$, for any $\lambda_0 > 0$,
\begin{align*}
& \lim_{\epsilon \to 0} P(\mbox{at least one individuals in } [t, t+\epsilon] \mbox{ with rate } \leq \lambda_0 \mid A) \\
& \leq \lim_{\epsilon \to 0} \frac{O(\epsilon)/\epsilon}{ 
(\psi(\epsilon)/\epsilon) \max \left\{ 1-\psi(\epsilon)/2, 0 \right\} } = 0.~~~~~~~~~~~~~~~~~~~~~~~~~\Box
\end{align*}

\section{Proof of the equivalence between Condition (\ref{eq:nu_finite}) and the finiteness of ESAC}\label{AppB} 

\noindent
Suppose Condition (\ref{eq:nu_finite}) holds. Then 
\[
\int\frac{1-e^{-\lambda t}}{\lambda}\nu(d\lambda) \le\int\min \left\{t,\frac{1}{\lambda} \right\}\nu(d\lambda) \le\max\{t,1\}\int\min \left\{1,\frac{1}{\lambda} \right\}\nu(d\lambda) < \infty
\]
for $t \geq 0$. On the other hand, if $E(N_+(1))<\infty$, then 
\[
\int \min\{1,\lambda^{-1}\}\nu(d\lambda) \leq \frac{1}{1-\exp(-1)} \int \frac{1-\exp(-\lambda )}{\lambda} \nu(d\lambda) = \frac{E(N_+(1))}{1-\exp(-1)} < \infty.
\]

\section{Equivalent definition of the MPPP}\label{AppA}

\noindent
In the definition of the MPPP in the paper, the individuals of the zero-rate species and the individuals of the positive-rate species are generated separately. Here we present an alternative definition where all individuals are generated in a unified manner.   

\vspace{0.3cm}
\noindent
{\bf Definition: }(Equivalent definition of MPPP)
\label{def:altdef} An MPPP $G$ is characterized by a nonzero species intensity
measure $\nu$, which is a measure over $\mathbb{R}_{\ge0}$ satisfying $\int_0^\infty \min\{1,\lambda^{-1}\} \nu(d \lambda)<\infty$. Define $\mathring{\nu}$
to be a measure over $\mathbb{R}_{\ge0}^{2}$ by
\[
\mathring{\nu}(A)=\int\int_{0}^{\infty}\mathbf{1}\{(\lambda,t)\in A\}e^{-\lambda t}dt\cdot\nu(d\lambda) 
\]
for any measurable set $A\subseteq \mathbb{R}_{\ge0}^{2}$. Generate
$(\lambda_{1},t_{1}),(\lambda_{2},t_{2}),\ldots$ (a finite or countably
infinite sequence) according to a Poisson process with intensity measure
$\mathring{\nu}$. For each simulated $(\lambda_i,t_i)$, we generate a realization $\eta_{i}$ (independently across
$i$) of a Poisson process with rate $\lambda_{i}$, conditioned
on the event that the first point is at time $t_{i}$. (That is, it
contains the point $t_{i}$ together with a Poisson process starting
at time $t_{i}$. If $\lambda_{i}=0$, then $\eta_{i}$ contains only 
one point $t_{i}$.) Each $\eta_i$ stores the appearance times of one species with rate $\lambda_i$. Finally, we take $G=\{\eta_{1},\eta_{2},\ldots\}$.

\vspace{0.3cm}
This definition models the species that will eventually
be observed in a study. The first appearance time for each of such
species (i.e., the first point of each $\eta_{i}$) is explicitly included
in the definition of $\mathring{\nu}$. 

\section{Proof of the Probability that an Individual Observed in a Given Future Time Belongs to a Species Represented $k$ times in $[0,t_0]$}\label{AppL}
Let $\lambda^*$ be the rate of the species observed in a given future time. Suppose $\Lambda < \infty$. From Proposition A in Appendix \ref{AppFix}, the probability measure of $\lambda^*$ is $\nu/\Lambda$. Thus
\begin{align*}
& P(\mbox{individual observed in a given future time  belongs to a species } \\
& \mbox{represented } k \mbox{ times in }[0,t_0]) \nonumber \\
& = E[P(\mbox{individual observed in a given future time belongs to a species } \\
& \mbox{represented } k \mbox{ times in }[0,t_0]~|~ \lambda^*)] \nonumber \\
& = E\left[\frac{(\lambda^* t_0)^k\exp(-\lambda^* t_0)}{k!} \right] \\
& = \frac{1}{\Lambda} \int \frac{(\lambda^* t_0)^k\exp(-\lambda^* t_0)}{k!} \nu(d \lambda^*) \\
& = \frac{(k+1)E(N_{k+1}(t_0))}{E(S(t_0))}.
\end{align*} 
The last expression holds also when $\Lambda = \infty$ (i.e., the above probability is zero) because from Proposition B in Appendix \ref{AppFix}, the rate of the future individual approaches infinity and thus that species should have been seen infinite number of times in $[0, t_0]$. 

\section{Proof of Equation (\ref{eq:sad1})} \label{appprob}
\begin{eqnarray*}
p_k(t) & = & \Pr(\mbox{a species is represented } k \mbox{ times in time interval } [0,t] \mid \\
& & \mbox{it is observed in time interval } [0,t] ) \\
& = & E(\Pr(\mbox{a species is observed } k \mbox{ times in time interval } [0,t] \mid \\
& & \mbox{the rate of the recorded species is } \lambda)~ \mid ~ \mbox{the species is observed in }  \\
& & \mbox{ time interval } [0,t] ) \\
& = & E\left( \left. \frac{(\lambda t)^k \exp(-\lambda t)/k!}{1-\exp(-\lambda t)}~ \right|~ \mbox{it is observed in time interval } [0,t]  \right).
\end{eqnarray*}
From the thinning property of Poisson processes, the probability measure for $\lambda$ given that the species is observed in time interval $[0,t]$  is
\[
\frac{\lambda^{-1}(1-\exp(-\lambda t)) \nu}{\int \lambda^{*-1} (1-\exp(-\lambda^* t)) \nu (d \lambda^*)}.
\]
Therefore,
\begin{eqnarray*}
p_k(t) & = & \int \frac{(\lambda t)^k \exp(-\lambda t)/k!}{1-\exp(-\lambda t)} \frac{\lambda^{-1}(1-\exp(-\lambda t)) }{\int \lambda^{*-1} (1-\exp(-\lambda^* t)) \nu (d \lambda^*)} \nu(d \lambda) \\
& = & \frac{\int \lambda^{k-1} t^k \exp(-\lambda t)/k! \nu(d \lambda)} {\int \lambda^{*-1} (1-\exp(-\lambda^* t)) \nu (d \lambda^*)} \\
& = & \frac{E(N_k(t))}{E(N_+(t))}.
\end{eqnarray*}

\section{Proof of Proposition 1}\label{AppProp1} 

\noindent
For any MPPP, from \eqref{eq:esac_dk}, $\psi(t)$ is a Bernstein function. Clearly $\psi(0)=0$. It proves the necessity part of (a). The sufficiency part of (a) follows from the fact that every Bernstein function $g(t)$ with $g(0)=0$
has a unique L\'evy-Khintchine representation
\[
g(t)=\kappa t+\int_{0}^{\infty}(1-\exp(-\lambda t))\mu(d\lambda),
\]
where $\kappa \ge 0$, and $\mu$ is a measure over $[0,\infty)$ such that
$\int_{0}^{\infty}\min\{1,\lambda\}\mu(d\lambda)<\infty$. 
Define an MPPP with $\nu(\{0\})=\kappa$ and $\tilde \nu = \mu$. The condition $\int_{0}^{\infty}\min\{1,\lambda\}\mu(d\lambda)<\infty$
becomes Condition \eqref{eq:nu_finite}. From (\ref{eq:En+}), the ESAC of this MPPP is equal to $g(t)$. It proves the  sufficiency part of (a). It also proves that $\psi(t)$ uniquely determines an MPPP in (c) because L\'evy-Khintchine representation is unique.  

To prove (d), we only need to show that $h_t(s)$ in (\ref{eq:pgf1}) is the probability generating function of $p(t)$. Clearly $h_t(0)=0$. From (\ref{eq:deriv-1}), for $k\geq 1$, $h_t^{(k)}(0)/k! = (-1)^{k+1}t^k \psi^{(k)}(t)/(k! \psi(t)) = E(N_k(t))/\psi(t)=p_k(t)$. It completes the proof of (d).

Given $h_{t_0}(s)$ and $\psi(t_0)$ for a fixed $t_0 > 0$, from (\ref{eq:pgf1}), $\psi(t) = \psi(t_0) (1-h_{t_0}(1-t/t_0))$. The ESAC is uniquely determined and so is the MPPP. It completes the proof for the remaining part of (c).

To prove the necessity of (b), let $g(s)$ be the probability generating function of $p(t_0)$. From (\ref{eq:pgf1}), 
\[
g(s) = 1 - \frac{\psi((1-s)t_0)}{\psi(t_0)}.
\]
Thus $g(0)=0$ and $g(1) = 1$. Furthermore,
\[
\psi(t) = \psi(t_0) (1-g(1-t/t_0)).
\]
It follows that when $k \geq 1$, $\psi^{(k)}(t) = (-1)^{k+1}\psi(t_0)g^{(k)}(1-t/t_0)/t_0^k$. Because $\psi^{(k)}(t)$ has sign $(-1)^{k+1}$, $g^{(k)}(s) \geq 0$ for $s \in (-\infty, 1)$. It completes the proof of the necessity part. For the sufficiency part, suppose $g(0)=0, g(1)=1$, $g(s)$ is absolutely monotone in $(-\infty,1)$. Let $t_0$ and $\psi(t_0)$ be two given positive values. Define $\chi(t) = \psi(t_0)(1-g(1-t/t_0))$. It can be shown that $\chi(t)$ is a Bernstein function such that $\chi(0)=0$. From (a), there is an MPPP with ESAC equal to $\chi(t)$. We have $\chi(t_0)=\psi(t_0)$. From (\ref{eq:pgf1}), $g(s) = h_{t_0}(s)$ is the probability generating function of $p(t_0)$ of this MPPP. It completes the proof of (b).

\section{Proof of Proposition 2} \label{AppC}

Let $\xi(t)$ satisfy conditions (i) and (ii) in Proposition 2. Therefore, there exist $\epsilon >0$ and $\delta >0$ such that for all $0 < t < \delta$, we have $\xi(t) > (1+\epsilon)t$. Let $g(t) = \delta \int_0^t \exp(\int_y^\delta (1/\xi(x)) dx) dy$. We show that $g(t)$ is a $\psi(t)$ in the Proposition.
 
When $0 < t \leq \delta$, 
\[
g(t) 
 \leq  \delta \int_0^\delta \exp \left(\int_y^\delta \frac{1}{(1+\epsilon)x} dx \right) dy = \delta \int_0^\delta \left(\frac{\delta}{y} \right)^{1/(1+\epsilon)} dy =  \delta^2 \frac{1+\epsilon}{\epsilon} < \infty.
\]
When $t > \delta$,
\begin{eqnarray*}
g(t) & = & \delta \int_0^\delta \exp \left(\int_y^\delta (1/\xi(x)) dx \right) dy + \delta \int_\delta^t \exp\left(- \int_\delta^y (1/\xi(x)) dx \right) dy \\
& \leq & \delta^2 (1+\epsilon)/\epsilon + \delta \int_\delta^t 1 dy < \infty.
\end{eqnarray*}
Therefore, $g(t)$ is finite for any $t>0$. 
As $g^{(1)}(t) = \delta \exp(\int_t^\delta (1/\xi(x)) dx)$, and $g^{(2)}(t) = -\delta \exp(\int_t^\delta (1/\xi(x)) dx) (1/\xi(t))$, we have $-g^{(1)}(t)/g^{(2)}(t) = \xi(t)$. Since $g(0)=0$, from Proposition 1(a), it is sufficient if we can prove that  
\begin{equation}
(-1)^{k-1} g^{(k)}(t) \geq 0,~~~(k \geq 1). \label{eq:ben}
\end{equation}
Clearly (\ref{eq:ben}) holds when $k=1$. Suppose (\ref{eq:ben}) holds when $k \leq d$ for a $d \geq 1$. As $g^{(2)}(t) \xi(t) = -g^{(1)}(t)$, after taking derivative $(d-1)$ times on both sides, we have
\[
\sum_{i=0}^{d-1} {d-1 \choose i} g^{(d+1-i)}(t) \xi^{(i)}(t) = -g^{(d)}(t).
\]
For $i \geq 1$, $sign(g^{(d+1-i)}(t) \xi^{(i)}(t)) = sign((-1)^{d-i}(-1)^{i+1}) = (-1)^{d+1}$. Furthermore, $sign(-g^{(d)}(t))= (-1)^d$. Thus $sign(g^{(d+1)}(t) \xi(t)) = (-1)^d$. It implies that $sign(g^{(d+1)}(t))=(-1)^d$ completing the proof of (\ref{eq:ben}) by induction.

\section{Proof of Proposition 3}\label{AppD}

\noindent
If an SAD is time invariant, then its probability generating function, $h_t(s)$ does not depend on $t$. From (\ref{eq:pgf1}), there is a positive function $g$ such that $\psi(ty) = \psi(t)g(y)$ for $y \in [0,\infty)$. 
Take logarithm on both sides, take derivative with respect to $t$, and then set $t=1$. We have
\[
\frac{d \log(\psi(y))}{dy} = \frac{\psi^{(1)}(1)}{\psi(1)y}.
\]
The solution of the above differential equation is $\psi(y)= \psi(1) y^{\psi^{(1)}(1)/\psi(1)}$. As $\psi(t)$ is a Bernstein function, $0 < \psi^{(1)}(1)/\psi(1) \leq 1$. Hence power law is the only law with $p(t)$ independent on $t$.

Consider the second part of Proposition 3. Suppose an SAD in MPPP converges to a proper distribution with probability generating function $f(s)$. Then the $h_t(s)$ of this SAD converges to $f(s)$ for any $s \in [0,1]$. Let $x$ and $y$ be any two values in $[0,1]$. From (\ref{eq:pgf1}), $1-f(1-xy)= \lim_{t \to \infty} \psi(xyt)/\psi(t) = \lim_{t\to \infty} \psi(xyt)/\psi(xt) \lim_{t \to \infty} \psi(xt)/\psi(t) = (1-f(1-y))(1-f(1-x))$. Write $\pi(x)=(1-f(1-x))$. Then $\pi(xy)=\pi(x)\pi(y)$. Similar to the proof above, we have $\pi(x) = x^\alpha$ when $x\in [0,1]$. It implies that $f(s) = 1-\pi(1-s) = 1-(1-s)^\alpha$. As $f(1)=1$, $\alpha > 0$. Since $P(1) = f^{(1)}(0) = \alpha$, we have $0 < \alpha \leq 1$. It completes the proof as the probability generating function of power law is $1-(1-s)^{1-1/c}$. 

\section{Proof of $\mathrm{LDR}_1 \subsetneq \mathrm{LDR}_2 = \mathrm{LDR}_3 = \ldots$}\label{AppE}

\noindent
For any positive integer $j$, condition $-\psi^{(j)}(t)=(b+ct)\psi^{(j+1)}(t)$ implies 
\[
-\psi^{(j+1)}(t)=c \psi^{(j+1)}(t) +(b+ct)\psi^{(j+2)}(t). 
\]
Therefore, $\mathrm{LDR}_j \subseteq \mathrm{LDR}_{j+1}$ for $j \geq 1$. 
Let $\phi(t) = \alpha t + \psi(t)$ where  $\alpha > 0$ and $\psi(t) \in \mathrm{LDR}_1$. Then $-\phi^{(2)}(t)/\phi^{(3)}(t) = -\psi^{(2)}(t)/\psi^{(3)}(t)$ which is a linear function of $t$ because $\psi(t) \in \mathrm{LDR}_1 \subseteq \mathrm{LDR}_2$. Clearly $\phi(t) \in \mathrm{LDR}_2$ but not in $\mathrm{LDR}_1$. Therefore, $\mathrm{LDR}_1 \neq \mathrm{LDR}_2$. It can be shown that every element in $\mathrm{LDR}_2$ has the form $\alpha t + \psi(t)$ for a $\psi(t) \in \mathrm{LDR}_1$. It means that $\mathrm{LDR}_2$ is a mixture of zero-rate species and $\mathrm{LDR}_1$.

Consider $\psi(t) \in \mathrm{LDR}_j$ for $j \geq 3$. Let  $-\psi^{(j)}(t)/\psi^{(j+1)}(t)=b+ct$ (i.e., $d \log(\psi^{(j)}(t))/dt = -1/(b+ct)$). As $j$th derivative ratio is always a nonnegative nondecreasing function of $t$, both $b$ and $c$ are nonnegative. For simplicity, we only consider the case when $c>0$ and $c \neq 1$ (cases when $c=0$ and $c=1$ can be studied through letting $c \to 0$ and $c \to 1$ respectively). Then 
\begin{equation}
\psi^{(j)}(t) = \psi^{(j)}(1)[(b+ct)/(b+c)]^{-1/c}, \label{eq:a1}
\end{equation} 
\begin{equation}
\psi^{(j-1)}(t)=\left[ \psi^{(j-1)}(1) -\psi^{(j)}(1) \left(\frac{b+c}{c-1}\right) \right] + \psi^{(j)}(1) \left(\frac{b+c}{c-1}\right)\left(\frac{b+ct}{b+c}\right)^{1-1/c} , \label{eq:a2}
\end{equation}
and 
\begin{eqnarray}
\psi^{(j-2)}(t) & = & \psi^{(j-2)}(1)+\left[\psi^{(j-1)}(1)-\psi^{(j)}(1)\left(\frac{b+c}{c-1}\right)\right](t-1) \nonumber \\
& & +\frac{(b+c)^2\psi^{(j)}(1)}{(c-1)(2c-1)} \left[\left(\frac{b+ct}{b+c}\right)^{2-1/c}-1\right]. \label{eq:a3}
\end{eqnarray}
If $c > 1$ and $\psi^{(j)}(1) \neq 0$, from (\ref{eq:a2}), $sign(\psi^{(j-1)}(t)) = sign(\psi^{(j)}(1)) \neq 0$ when $t$ is large. It is impossible because they should have different sign. If $c > 1$ and $\psi^{(j)}(1)=0$, from (\ref{eq:a1}), $\psi^{(j)}(t)$ is a zero function. It is impossible as $-\psi^{(j)}(t)/\psi^{(j+1)}(t)$ is undefined. If $0 < c < 1$, from (\ref{eq:a2}) and (\ref{eq:a3}), $sign(\psi^{(j-1)}(t)) = sign(\psi^{(j-1)}(1) + \psi^{(j)}(1)[(b+c)/(1-c)]) = sign(\psi^{(j-2)}(t))$ when $t$ is large. It is possible only when $sign(\psi^{(j-1)}(t))=0$. It implies that $\psi^{(i-1)}(1) + \psi^{(i)}(1)[(b+c)/(1-c)] = 0$. From (\ref{eq:a1}) and (\ref{eq:a2}),  $-\psi^{(j-1)}(t)/\psi^{(j)}(t) = (b+ct)/(1-c)$. Thus $\psi(t) \in \mathrm{LDR}_{j-1}$. It follows that $\mathrm{LDR}_j = \mathrm{LDR}_{j-1}$ for all $j \geq 3$. 

\section{Pointwise Confidence Band for D2/D3 plot}\label{AppF}

\noindent
Similar to the confidence band for D1/D2 plot, an approximate 95\% pointwise confidence band for D2/D3 plot is $-\hat \psi^{(2)}(t)/\hat \psi^{(3)}(t) \pm 1.96 \sqrt{\widehat{Var}(-\hat \psi^{(2)}(t)/\hat \psi^{(3)}(t))}$, where
\begin{align*}
& \widehat{Var}\left(-\frac{\hat \psi^{(2)}(t)}{\hat \psi^{(3)}(t)}\right) \\
& = \frac{1}{\hat \psi^{(3)2}(t)} \widehat{Var}(\hat \psi^{(2)}(t)) + \frac{\hat \psi^{(2)2}(t)}{\hat \psi^{(3)4}(t)}\widehat{Var}(\hat \psi^{(3)}(t)) - \frac{2\hat \psi^{(2)}(t)}{\hat \psi^{(3)3}(t)}\widehat{Cov}(\hat \psi^{(2)}(t),\hat \psi^{(3)}(t)),
\end{align*}
with
\[
\widehat{Var}(\hat \psi^{(2)}(t)) = \frac{1}{t_0^4} \sum_{k=2}^\infty k^2(k-1)^2 N_k(t_0) \left(1- \frac{t}{t_0} \right)^{2k-4},
\]
\[
\widehat{Var}(\hat \psi^{(3)}(t)) = \frac{1}{t_0^6} \sum_{k=3}^\infty k^2(k-1)^2(k-2)^2 N_k(t_0) \left(1- \frac{t}{t_0} \right)^{2k-6},
\]
and 
\[
\widehat{Cov}(\hat \psi^{(2)}(t),\hat \psi^{(3)}(t)) = -\frac{1}{t_0^5} \sum_{k=3}^\infty k^2(k-1)^2 (k-2) N_k(t_0) \left(1- \frac{t}{t_0} \right)^{2k-5}.
\]

\section{Parametric Bootstrap Confidence Interval for Hill Numbers} \label{Appboot}
We have $\hat \theta$, an estimator of $\theta$ which is $E(N_0(t_0))$ or a Hill number of order $q$. We want to construct a $100(1-\alpha)\%$ confidence interval for $\theta$. Let $B$ be the total number of bootstrap samples such that $\alpha(B+1)/2$ is an integer. In this paper, we use $\alpha = 0.05$ and $B=2999$. As $\hat \theta$ can be infinite, we avoid using methods that require arithmetic on $\hat \theta$, such as the basic method. The procedure to construct a parametric bootstrap confidence interval is as follows:

\vspace{0.3cm}
\noindent
For $i = 1$ to $B$ do \{ 

\noindent
~~~~~Generate a bootstrap sample from a parametric model.

\noindent
~~~~~Compute $\hat \theta$, which we denote as $\hat \theta_i$ basing on the bootstrap sample. \}   

\noindent
Sort $\hat \theta_1, \hat \theta_2, \ldots, \hat \theta_B$ in ascending order. Denote the sorted $\hat \theta_1, \hat \theta_2, \ldots, \hat \theta_B$ as $\hat \theta_{(1)} \leq \hat \theta_{(2)} \leq \ldots \leq \hat \theta_{(B)}$. Set $\hat \theta_{(0)}=0$ and $\hat \theta_{(B+1)}=\infty$. A $100(1-\alpha)\%$ confidence interval for $\theta$ is
\[
[~ \hat \theta_{(j)}~,~ \hat \theta_{((B+1)(1-\alpha)+j)}~].
\]
where $j$ is an integer such that $0 \leq j \leq (B+1)\alpha$ and the above $100(1-\alpha)\%$ confidence interval is smallest. 

\vspace{0.3cm}
Infinity is a special value of $\hat \theta$ which can have positive probability mass. We use the smallest $100(1-\alpha)\%$ confidence interval, and attempt to construct confidence interval with finite upper endpoint if possible.

When $\hat \theta$ is an estimator of a Hill number under parametric model $\mathrm{LDR}_1$ or $\mathrm{RDR}_1$, the bootstrap sample is a simulated FoF under the selected model with the parameter equal to the MLE.  This simulation consists of two steps: (i) simulate $N_+(t_0)$ from $\mathrm{Poisson}(\hat E(N_+(t_0)))$ distribution, and (ii) generate a random sample  of size $N_+(t_0)$ from the fitted SAD and they form the FoF.

When $\hat \theta=\hat E(N_0(t_0))$ in (\ref{eq:unseen}), the bootstrap sample is $\{N_i^*(t_0)\}_{i=1,2,3}$. We simulate $N_i^*(t_0)$ from $\mathrm{Poisson}(N_i(t_0))$ distribution for $i=1,2,3$ independently across $i$. After a confidence interval for $E(N_0(t_0))$ is constructed, add $N_+(t_0)$ to it to find a confidence interval for $E(D)$. If at least one of $N_1(t_0), N_2(t_0)$ or $N_3(t_0)$ is zero, modification is recommended to avoid having any $N_i^*(t_0)$ ($i=1,2,3$) to be fixed to 0. When there is $j \geq 3$ such that $N_j(t_0)>0$, we move the ending time $t_0$ a little bit backward to $t$, say $t = t_0 - t_0/S(t_0)$ where $S(t_0)$ is the total number of individuals observed in time interval $[0,t_0]$, and find $\{\hat N_i(t)\}_{i=1,2,3}$ using (\ref{eq:nj}) as shown below: 
\[
\hat N_i (t) = \sum_{k=i}^\infty N_k(t_0) {k \choose i} \left( \frac{t}{t_0} \right)^i \left(1 - \frac{t}{t_0} \right)^{k-i}.
\]
Clearly $\hat N_i(t) > 0$ for $i=1,2,3$. We use $\{\hat N_i(t)\}_{i=1,2,3}$ in place of $\{N_i(t_0)\}_{i=1,2,3}$ in bootstrapping. We estimate $N_+(t)$ ($=\psi(t)$) using the relation in (\ref{eq:psiv}). Add this estimate to the confidence interval for $N_0(t)$ to construct a confidence interval for $E(D)$.

\section{Expressions of $^q\!D_\nu$ for $\mathrm{RDR}_1$} \label{HillRDR}

\noindent
If $c_1$ or $c_2$ is zero, $\mathrm{RDR}_1$ reduces to $\mathrm{LDR}_1$. Assume that $c_1$ and $c_2$ are positive.
\begin{eqnarray*}
\frac{d \tilde \nu}{d \lambda} & = & \frac{a(t_0+b_1)^{c_1}(t_0+b_2)^{c_2} \exp(-b_1 \lambda) \lambda^{c_1+c_2-2}}{\Gamma(c_1)\Gamma(c_2)} \\
& & \int_0^1 y^{c_2-1}(1-y)^{c_1-1}\exp(-(b_2-b_1)\lambda y) dy.
\end{eqnarray*}
Assume further that $b_1 > 0$, which implies that $\Lambda$ is finite. When $q \geq 0$ and $q \neq 1$, 
\begin{eqnarray*}
^q\!D_\nu & = & a \left(\frac{t_0+b_1}{b_1} \right)^{c_1} \left(\frac{t_0+b_2}{b_2} \right)^{c_2}  \left(\frac{b_1b_2 \Gamma(c_1+c_2+q-1)}{\Gamma(c_1)\Gamma(c_2)} \right. \\
& & \left. \int_0^1 \vartheta(y)^{c_2-1}(1-\vartheta(y))^{c_1-1}((b_2-b_1)y+b_1)^{-q-1} dy \right)^{1/(1-q)},
\end{eqnarray*}
where $\vartheta(y) = b_2y/(b_2y + b_1(1-y))$. 
When $q=1$,
\begin{eqnarray*}
^1\!D_\nu & = & a \left(\frac{t_0+b_1}{b_1} \right)^{c_1} \left(\frac{t_0+b_2}{b_2} \right)^{c_2} \exp \left( -\frac{ b_1 b_2}{ B(c_1,c_2) } \right. \\
& & \left. \int_0^1 \vartheta(y)^{c_2-1} (1-\vartheta(y))^{c_1-1} \frac{\Psi(c_1+c_2) - \log((b_2-b_1)y+b_1)} {((b_2-b_1)y+b_1)^{2}} dy \right),
\end{eqnarray*}
where $B(x,y)$ is the Beta function and $\Psi(x)$ is the digamma function.

\section{Proof of the log-likelihood function for $\rho$-appearance design}\label{AppH}

\noindent
Let $Y_i$ be the length of the observation period for species $i$, and $J_i$ be the observed frequency of species $i$ in its observation period. For species $i$, we observe $(Y_i,J_i)$. Either $Y_i=t_0$ or $J_i=\rho$. For a species with rate $\lambda$, the probability function of $J$ is 
\begin{equation}
P(J=j \mid \lambda)= \begin{cases}
(\lambda t_{0})^{j}\exp(-\lambda t_{0})/j! & (j<\rho) \\
\sum_{k=\rho}^{\infty} (\lambda t_{0})^{k} \exp(-\lambda t_{0})/k! & (j=\rho).
\end{cases}\label{eq:inc_pj-1}
\end{equation}
Since the time of the $\rho$th individual follows $\mathrm{Erlang}(\rho,\lambda)$ distribution,
\begin{align}
P(Y\in[y,y+dy),\,J=\rho\, \mid \,\lambda) & =\frac{\lambda^{\rho}y^{\rho-1}\exp(-\lambda y)}{(\rho-1)!}dy.\label{eq:inc_pt-1}
\end{align}
By \eqref{eq:inc_pj-1} and \eqref{eq:inc_pt-1}, the joint probability
density function of the observations $\{(Y_{i},J_{i})\}_{i\leq n_{+}(t_{0})}$, say $\{(y_{i},j_{i})\}_{i\leq n_{+}(t_{0})}$
given $\nu$ is proportional to 
\[
e^{-\psi(t_{0})}\left(\prod_{i:\,j_{i}<\rho}\int\frac{\lambda^{j_{i}-1}t_{0}^{j_{i}}e^{-\lambda t_{0}}}{j_{i}!}\nu(d\lambda)\right)\left(\prod_{i:\,j_{i}=\rho}\int\frac{\lambda^{\rho-1}y_{i}^{\rho-1}e^{-\lambda y_{i}}}{(\rho-1)!}\nu(d\lambda)\right).
\]
Therefore, the log-likelihood function is 
\[
\log(\mathcal{L}(\psi \mid \{y_i,j_i\}_i)) 
= -\psi(t_{0})+\sum_{j=1}^{\rho-1} n_j(t_0) \log(\mid \psi^{(j)}(t_{0}) \mid ) + \sum_{y_{i}<t_{0}} \log(\mid \psi^{(\rho)}(y_{i}) \mid ).
\]

\section{Simulation experiment on the loss of information of the $\rho$-appearance design}\label{AppI}

\vspace{0.3cm}
\noindent
Let us reconsider the bird abundance data for the Wisconsin route of the North
American Breeding Bird Survey for 1995 in Section \ref{s:swine}. We choose 
$\nu(d\lambda)=\gamma \lambda f(\lambda \mid \mu,\sigma)d\lambda$
where $f(\lambda \mid \mu,\sigma)$ is the density function of $\mathrm{Lognormal}(\mu,\sigma^{2})$
distribution.} The parameter $\gamma$ is the expected total number
of species. From Section \ref{s:FoF}, the MLE of $(\mu, \sigma)$ is identical
to the conditional maximum likelihood estimate of the corresponding Poisson-lognormal model \citep{bulmer1974}.
The fitted lognormal mixing distribution is $\mathrm{Lognormal}(\hat{\mu},\hat{\sigma}^{2})$
distribution with $\hat{\mu}=1.23$ and $\hat{\sigma}=1.30$.
Let $\omega_{i}(\mu,\sigma)=P(Y=i)$ where $Y$ is a Poisson-lognormal
random variable with parameters $\mu$ and $\sigma$. We use the function
``dpoilog'' in R-package ``poilog'' \citep{grotan2015} to compute this
probability. The estimated $\gamma$ is $\hat{\gamma}=n_{+}(t_{0})/(1-\omega_{0}(\hat{\mu},\hat{\sigma}^{2}))=85.2$.

Without loss of generality, set $t_{0}=1$. We use this data to investigate
the information loss of the $\rho$-appearance design. The $\rho$-appearance
data are simulated from the data using the following procedure:

\vspace{0.3cm}
\noindent
{\em Simulation procedure:} Suppose species $i$ has observed frequency
$m_{i}$ in time $[0,1].$ If $m_{i}<\rho$, our data for this species
is $m_{i}$, the frequency of it in time $[0,1].$ If $m_{i}\geq\rho$,
we simulate the $\rho$-appearance time of the species, $r_{i}$ from
$\mathrm{Beta}(\rho,m_{i}+1-\rho)$ distribution, which is the distribution
of the $\rho$ order statistic of $m_{i}$ samples from the $U(0,1)$
distribution.

\begin{figure}
\begin{center}

\includegraphics[width=13cm,height=10cm]{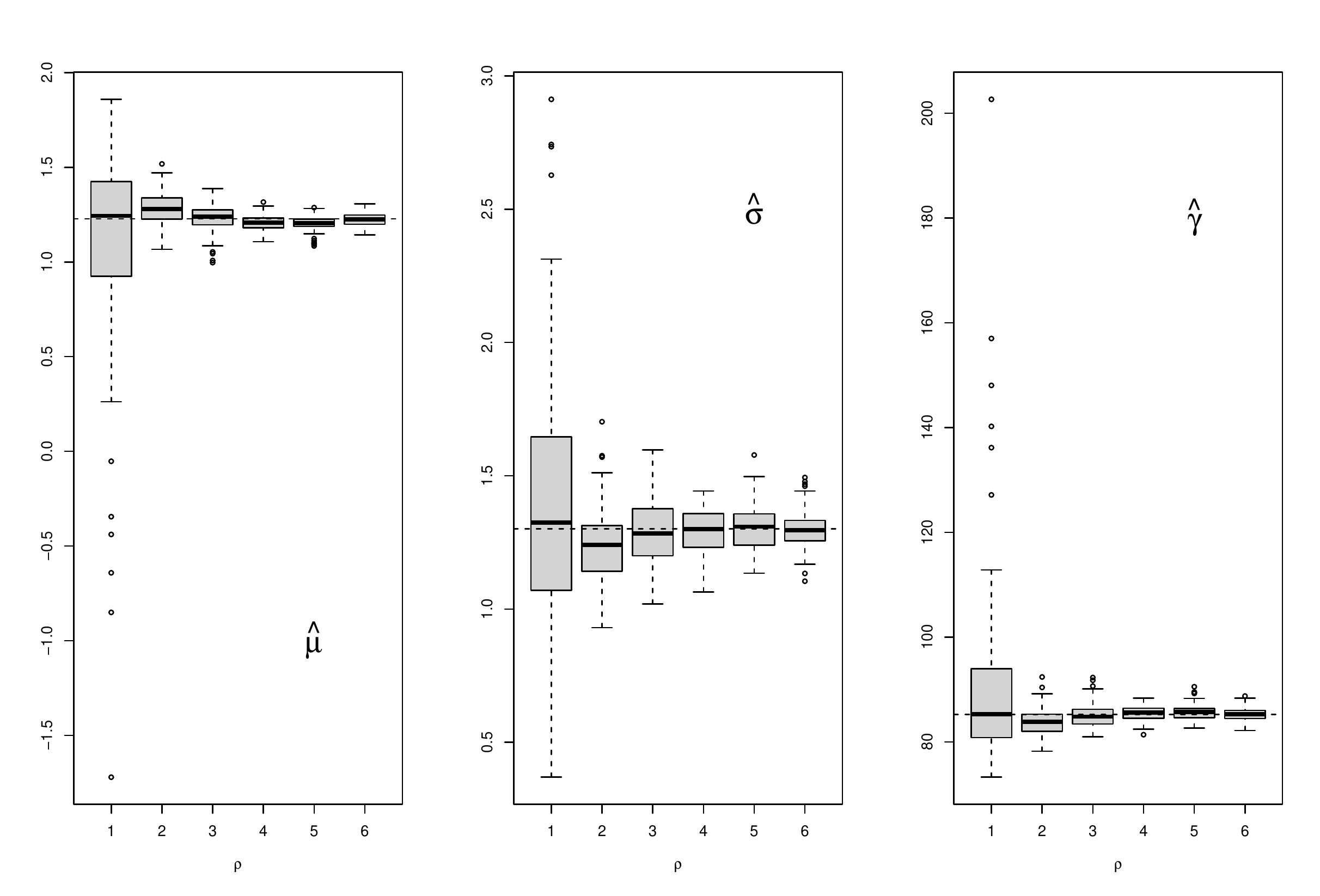}
\caption{Box-and-whisker plots for estimators across different values of $\rho$
in the simulation. The horizontal dashed
line in each plot shows the estimate when the full Wisconsin route of the North American breeding bird survey
data for 1995 is used. The variability of the estimate delineates the
additional noise due to the $\rho$-appearance design.}
\label{Boxplot}
\end{center}
\end{figure}

\vspace{0.3cm}
It can be shown that $E(N_{k}(t))=\gamma \omega_{k}(\mu+\log(t),\sigma)$. The log-likelihood function is
\begin{align*}
& \log(\mathcal{L}(\gamma,\mu,\sigma \mid \{n_j(1)\}_{j=1,\ldots,\rho-1},\{r_i\})) \\
& = -\gamma(1-\omega_{0}(\mu,\sigma))+\sum_{k=1}^{\rho-1}n_{k}(1)\log(\gamma \omega_{k}(\mu,\sigma))+\sum_{r_{i}}\log(\gamma \omega_{\rho}(\mu+\log(r_{i}),\sigma)).
\end{align*}
As the MLE of $E(N_{+}(1))=\gamma(1-\omega_{0}(\mu,\sigma))$ is $n_{+}(1)$,
MLE of $\mu$ and $\sigma$, say $\hat \mu$ and $\hat \sigma$ respectively can be found through maximizing the following function.
\[
-n_{+}(1)\log(1-\omega_{0}(\mu,\sigma))+\sum_{k=1}^{\rho-1}n_{k}(1)\log(\omega_{k}(\mu,\sigma))+\sum_{r_{i}}\log(\omega_{\rho}(\mu+\log(r_{i}),\sigma)).
\]
The MLE of $\gamma$ is $\hat{\gamma}=n_{+}(1)/(1-\omega_{0}(\hat{\mu},\hat{\sigma}))$.
We consider $\rho=1,2,...,6$. For each $\rho$-value, we simulate
100 independent sets of $\rho$-appearance data. For each simulated
data, $\mu$, $\sigma$ and $\gamma$ are estimated. The sample mean
and sample standard deviation of the estimates are presented
in Table \ref{A}. Graphical display is given in Figure~\ref{Boxplot}.

The mean of the estimate is close to that basing on $\tilde n(1)$. The standard deviation
of the estimator decreases as $\rho$ increases. From the simulation results, the standard deviation of the estimators when $\rho=1$ is considerably worse than those when $\rho=2$. 
Value $\rho=4$ performs well for this
data. The total number of species with frequency less than 4 is 33,
around 46\% of the seen species. 

\begin{table}
\begin{center}
\def~{\hphantom{0}}
\caption{Mean and standard deviation of MLE for North American breeding bird survey data (1995)}

\begin{tabular}{cccccccc}
\hline 
$\rho$ & 1 & 2 & 3 & 4 & 5 & 6 & $\infty$ \tabularnewline
\hline 
mean of $\hat{\mu}$ & 1.10 & 1.28 & 1.23 & 1.21 & 1.21 & 1.22 & 1.23\tabularnewline
sd of $\hat{\mu}$ & 0.56 & 0.09 & 0.07 & 0.04 & 0.04 & 0.03 & 0{*}\tabularnewline
\hline 
mean of $\hat{\sigma}$ & 1.39 & 1.24 & 1.29 & 1.29 & 1.31 & 1.30 & 1.30\tabularnewline
sd of $\hat{\sigma}$ & 0.48 & 0.15 & 0.13 & 0.08 & 0.09 & 0.07 & 0{*}\tabularnewline
\hline 
mean of $\hat{\gamma}$ & 90.4 & 83.8 & 85.0 & 85.5 & 85.7 & 85.3 & 85.2\tabularnewline
sd of $\hat{\gamma}$ & 18.7 & 2.52 & 2.32 & 1.38 & 1.56 & 1.29 & 0{*}\tabularnewline
\hline 
\end{tabular}

\label{A}
* When $\rho=\infty$, we always observe the full data, and the sample 

standard deviation (sd) of the estimator across simulation is zero.
\end{center}
\end{table}

\section{Simulation comparison of MLE method and curve-fitting method in extrapolation when finite number of points in the empirical SAC are available}\label{AppJ}

\noindent
Without loss of generality, we set $t_0=1$. Our data is $\{n_+(0.1), n_+(0.2), \ldots, n_+(1)\}$. We consider three distributions: power law, log-series distribution and geometric distribution. The curve-fitting methods for the distributions are described below:

\begin{description}

\item[] 
(a) Power law: $\psi(t)=\tau t^{1-1/c}$. For curve fitting, we regress $\log(n_+(t))$ on $\log(t)$.

\item[]
(b) Log-series distribution: $\psi(t) = \tau \log(1+t/b)/\log(1+1/b)$. For curve fitting, we regress $n_+(t)$ on $\log(t)$. It is the standard approximation method which assumes that $1+t/b \approx t/b$.

\item[]
(c) Geometric distribution (hyperbola law): $\psi(t) = \tau (1+2b)t/(t+2b)$. There are various curve-fitting methods for hyperbola law (see for example \citet{raaijmakers1998sta}). In this simulation, we regresses $1/n_+(t)$ on $1/t$.

\end{description}

The parameter $\tau = \psi(1)$ is the expected number of recorded species at time $t_0=1$. In the experiment, $\tau$ can take value 200 and 1000. The distribution parameter can take 6 values. For power law, the value of $c$ can be 1.25, 1.5, 2, 3, 4 and 5. For log-series distribution, the value of $b$ can be 0.01, 0.02, 0.03, 0.05, 0.1, and 0.2. For geometric distribution, the value of $b$ can be 0.05, 0.1, 0.2, 0.4, 0.6 and 0.8.  
For each combination of parameters, we simulate 5000 data. The MLE and the curve-fitting estimate of $\psi(2)$ and $\psi(4)$ are found for each simulated data. We evaluate the performance of an estimator by the root mean squared relative error (RMSRE) (for estimator $\hat \theta$ for $\theta$, $\mathrm{RMSRE} = \sqrt{\sum_{i=1}^n ((\hat \theta_i -\theta)/\theta)^2/n}$). The results of the simulation are presented in Tables \ref{B}, \ref{C} and \ref{D}. The RMSRE for MLE is smaller than that for the curve-fitting method in this simulation study.

\vspace{0.3cm}
\begin{table}
\begin{center}
\caption{Simulation results for Power law in extrapolation to $t$ for fixed $\psi(1)$ given ten equally spaced observations of SAC in time interval $[0,1]$}

\vspace{0.3cm}
\begin{tabular}{llcccccc}
\hline
($t,\psi(1)$) & & $c=1.25$ & $c=1.5$ & $c=2$ & $c=3$ & $c=4$ & $c=5$ \\ 
\hline
(2,200) & RMSRE(Curve-fitting) & 0.075 & 0.078 & 0.083 & 0.091 & 0.096 & 0.100 \\
& RMSRE(MLE) & 0.073 & 0.074 & 0.077 & 0.080 & 0.082 & 0.083 \\
\hline
(2,1000) & RMSRE(Curve-fitting) & 0.034 & 0.035 & 0.037 & 0.040 & 0.042 & 0.044 \\
& RMSRE(MLE) & 0.033 & 0.034 & 0.035 & 0.036 & 0.037 & 0.037 \\
\hline
(4,200) & RMSRE(Curve-fitting) & 0.081 & 0.089 & 0.103 & 0.120 & 0.132 & 0.140 \\
& RMSRE(MLE) & 0.078 & 0.084 & 0.093 & 0.103 & 0.109 & 0.112 \\
\hline
(4,1000) & RMSRE(Curve-fitting) & 0.036 & 0.040 & 0.045 & 0.052 & 0.057 & 0.060 \\
& RMSRE(MLE) & 0.035 & 0.038 & 0.042 & 0.046 & 0.049 & 0.050 \\
\hline
\end{tabular}
\label{B}
\end{center}
\end{table}

\begin{table}
\begin{center}
\caption{Simulation results for log-series distribution in extrapolation to $t$ for fixed $\psi(1)$ given ten equally spaced observations of SAC in time interval $[0,1]$}

\vspace{0.3cm}
\begin{tabular}{llcccccc}
\hline
($t,\psi(1)$) & & $b=.01$ & $b=.02$ & $b=.03$ & $b=.05$ & $b=.1$ & $b=.2$ \\ 
\hline
(2,200) & RMSRE(Curve-fitting) & 0.072 & 0.073 & 0.073 & 0.077 & 0.090 & 0.122 \\
& RMSRE(MLE) & 0.072 & 0.072 & 0.072 & 0.072 & 0.073 & 0.076 \\
\hline
(2,1000) & RMSRE(Curve-fitting) & 0.033 & 0.034 & 0.036 & 0.043 & 0.065 & 0.106 \\
& RMSRE(MLE) & 0.032 & 0.032 & 0.032 & 0.033 & 0.033 & 0.034 \\
\hline
(4,200) & RMSRE(Curve-fitting) & 0.074 & 0.075 & 0.077 & 0.084 & 0.110 & 0.166 \\
& RMSRE(MLE) & 0.073 & 0.074 & 0.074 & 0.076 & 0.079 & 0.086 \\
\hline
(4,1000) & RMSRE(Curve-fitting) & 0.034 & 0.037 & 0.042 & 0.055 & 0.091 & 0.155 \\
& RMSRE(MLE) & 0.033 & 0.033 & 0.034 & 0.034 & 0.035 & 0.038 \\
\hline
\end{tabular}
\label{C}
\end{center}
\end{table}

\begin{table}
\begin{center}
\caption{Simulation results for geometric distribution in extrapolation to $t$ for fixed $\psi(1)$ given ten equally spaced observations of SAC in time interval $[0,1]$}

\vspace{0.3cm}
\begin{tabular}{llcccccc}
\hline
($t,\psi(1)$) & & $b=.05$ & $b=.1$ & $b=.2$ & $b=.4$ & $b=.6$ & $b=.8$ \\ 
\hline
(2,200) & RMSRE(Curve-fitting) & 0.322 & 0.458 & 0.586 & 0.687 & 0.731 & 0.756 \\
& RMSRE(MLE) & 0.085 & 0.106 & 0.136 & 0.165 & 0.174 & 0.174 \\
\hline
(2,1000) & RMSRE(Curve-fitting) & 0.316 & 0.454 & 0.583 & 0.684 & 0.728 & 0.753 \\
& RMSRE(MLE) & 0.054 & 0.079 & 0.111 & 0.134 & 0.136 & 0.132 \\
\hline
(4,200) & RMSRE(Curve-fitting) & 0.320 & 0.462 & 0.601 & 0.716 & 0.768 & 0.798 \\
& RMSRE(MLE) & 0.102 & 0.145 & 0.218 & 0.311 & 0.366 & 0.397 \\
\hline
(4,1000) & RMSRE(Curve-fitting) & 0.315 & 0.458 & 0.598 & 0.713 & 0.766 & 0.796 \\
& RMSRE(MLE) & 0.075 & 0.122 & 0.188 & 0.254 & 0.279 & 0.287 \\
\hline
\hline
\end{tabular}
\label{D}
\end{center}
\end{table}

\end{document}